\documentclass[12pt, a4paper]{article}
\pdfoutput=1
\usepackage{graphicx}
\usepackage{amssymb}
\usepackage{amsmath}
\usepackage{bm}
\usepackage[table]{xcolor} 
\usepackage{cite}
\usepackage{slashed}
\usepackage{subfigure}
\usepackage{epstopdf}            
\usepackage{epsfig}
\usepackage{here}
\usepackage{comment}
\usepackage{booktabs} 
\usepackage{colortbl} 
\usepackage{wrapfig}
\usepackage{ascmac}
\usepackage{fancybox}  

\renewcommand{\thefootnote}{\fnsymbol{footnote}}
\def\beq#1\eeq{\begin{align}#1\end{align}}

\newcommand{\br}{\mathrm{BR}}
\newcommand{\pb}{\mathrm{pb}}
\newcommand{\GeV} {\,\text{GeV}}
\newcommand{\TeV} {\,\text{TeV}}
\newcommand{\eg}{{\em e.g.}}
\newcommand{\ie}{{\em i.e.}}

\renewcommand{\arraystretch}{1.3}

\RequirePackage{xspace}
\def\Bbar    {\kern 0.18em\overline{\kern -0.18em B}{}\xspace}

\usepackage{caption}
\definecolor{BlueViolet}{rgb}{0.2, 0.00, 0.7}
\definecolor{Blue}{rgb}{0.15, 0.00, 0.9}
\definecolor{lightblue}{rgb}{0.15, 0.35, 0.95}
\definecolor{kitgreen}{rgb}{0, 
0.58823 
, 0.50980 
}
\usepackage[
colorlinks=true, linkcolor=lightblue,citecolor=lightblue,urlcolor=kitgreen]{hyperref}

\begin{document}
\sloppy 

\begin{titlepage}

\begin{center}

\hfill{P3H--22--047, TTP22--027, KEK--TH--2424}

\vskip .4in

\makeatletter\g@addto@macro\bfseries{\boldmath}\makeatother

{\Large{\bf Scrutinizing the 95--100 GeV di-tau excess\\[0.3em]
in the top associated process}}

\vskip .4in

{\large 
Syuhei Iguro$^{\rm (a,b)}$,
}
{\large 
Teppei Kitahara$^{\rm (c,d,e)}$,
 and} 
{\large
Yuji Omura$^{\rm (f)}$}
\vskip .2in

{\small
\begin{tabbing}
$^{\rm (a)}$ \= {\it 
Institute for Theoretical Particle Physics (TTP), Karlsruhe Institute of Technology (KIT),}\\[0.3em]
\> {\it  Engesserstra{\ss}e 7, 76131 Karlsruhe, Germany}
\\[0.3em]
$^{\rm (b)}$ \> {\it Institute for Astroparticle Physics (IAP),
KIT, 
Hermann-von-Helmholtz-Platz 1,}\\[0.3em]
\> {\it 76344 Eggenstein-Leopoldshafen, Germany}
\\[0.3em]
$^{\rm (c)}$ \> {\it 
Institute for Advanced Research, Nagoya University, Nagoya 464--8601, Japan}
\\[0.3em]
$^{\rm (d)}$ \> {\it 
Kobayashi-Maskawa Institute for the Origin of Particles and the Universe,} \\[0.3em]
\> {\it  Nagoya University, Nagoya 464--8602, Japan}
\\[0.3em]
$^{\rm (e)}$ \> {\it Theory Center, IPNS, High Energy Accelerator Research Organization (KEK), 1-1 Oho,} \\[0.3em]
\> {\it  Tsukuba 305--0801, Japan}
\\[0.3em]
$^{\rm (f)}$ \> {\it Department of Physics, Kindai University, Higashi-Osaka, Osaka 577--8502, Japan}
\\[0.2in]
%
\> {\it E-mail:} \href{mailto:igurosyuhei@gmail.com}{igurosyuhei@gmail.com}, \href{mailto:teppeik@kmi.nagoya-u.ac.jp}{teppeik@kmi.nagoya-u.ac.jp},\\[0.3em] \> \href{mailto:yomura@phys.kindai.ac.jp}{yomura@phys.kindai.ac.jp}
\end{tabbing}
}

\end{center}
\vskip .1in

\begin{abstract}
\noindent
Recently, the CMS collaboration has reported 
a di-tau excess with a local significance of 2.6--3.1$\sigma$
where the invariant mass is $m_{\tau\tau} =95$--100\,GeV.
This excess can be interpreted as a light scalar boson that 
couples to the third generation fermions, particularly top and $\tau$.
Based on the simplest model that can account for the CMS di-tau excess, 
we evaluate experimental sensitivities to the additional light resonance, using the results reported by the ATLAS collaboration. 
We see that a search for the top-quark associated production of the SM Higgs boson that decays into $\tau\bar\tau$ sets a strong model-independent limit.
We also find that the CP-even scalar interpretation of the light resonance is excluded by the ATLAS results, while the CP-odd interpretation is not.
\end{abstract}
\end{titlepage}

\setcounter{page}{1}
\renewcommand{\thefootnote}{\#\arabic{footnote}}
\setcounter{footnote}{0}

\hrule
\tableofcontents
\vskip .2in
\hrule
\vskip .4in


\section{Introduction}
\label{sec:intro}

The Standard Model (SM) has been experimentally verified with high accuracy and very successful. In the SM, the Higgs field plays a role in the mass generation of particles.
The potential for the Higgs field is given to break the electroweak (EW) symmetry, and particle masses are originated from the non-vanishing vacuum expectation value (VEV).
This mechanism has been tested by the EW precision observables, the 125-GeV Higgs measurements, and so on, and the predictions correspond reasonably well with the experimental results.
This success, however, poses a question about the origin of the Higgs potential. In particular, the mass squared of the Higgs field is negative and requires a severe fine-tuning to cancel the very large radiative corrections. In order to avoid the fine-tuning,
many models beyond the SM have been proposed so far; \eg, supersymmetry, composite Higgs, little Higgs, top partner,  extra dimensions, and gauge-Higgs unification.
These extended models generally predict additional particles around TeV scale, that can be tested in the experiments at the large hadron collider (LHC). Among the new particles, additional scalar fields
are often suggested as good candidates to validate the models.
The additional scalars, therefore, have been studied widely in both model-dependent and model-independent ways.

Interestingly, such a new particle, that interacts with the SM particles through the weak interaction and/or Yukawa interactions, can still take mass of $\mathcal{O}(100)\GeV$~\cite{ATLAS:2021moa,ATLAS:2022rme}, if the couplings with light quarks and leptons are suppressed.
The possibility of an additional scalar boson lighter than the 125\GeV~Higgs still remains.
Recently, 
the CMS collaboration has reported a new excess in the di-tau final states within all the $\tau$ decay modes (leptonic and hadronic) by using the Run 2 full data \cite{CMS:2022rbd},
which is an extension of the previous searches \cite{CMS:2011lzj,ATLAS:2012ube,ATLAS:2014vhc,ATLAS:2016ivh,ATLAS:2017eiz,CMS:2018rmh}. The excess can be interpreted as a resonance of a new particle. 
The local and global significance of the excess are $2.6\sigma $ and $2.3\sigma$ at the invariant mass of $m_{\tau\tau}=95\GeV$, respectively.\footnote{%
They have also reported a $2.8\sigma$ (local) and $2.4\sigma$ (global) excess at $m_{\tau\tau}=1.2\TeV$. It is, however, excluded by the same search by the ATLAS collaboration \cite{ATLAS:2020zms}.
} At $m_{\tau\tau}=100\GeV$, these values are 
 $3.1\sigma$ and $2.7\sigma$.
 By introducing an additional single neutral narrow resonance $\phi$, 
the best fit values for the excess are 
\cite{CMS:2022rbd}
\begin{align}
\sigma(gg \to \phi)\times \br(\phi \to \tau \bar\tau) =  7.7^{+3.9}_{-3.1}~\pb\quad \text{for~}m_{\phi} = 95\GeV\,,
\label{eq:tata95}
\end{align}
 or
\begin{align}
\sigma(gg \to \phi)\times \br(\phi \to \tau \bar\tau) =  5.8^{+2.4}_{-2.0}~\pb\quad \text{for~}m_{\phi} = 100\GeV\,.
\label{eq:tata100}
\end{align}
These cross sections are comparable to the $125\GeV$ SM Higgs boson, $h$ \cite{Anastasiou:2016cez},
\begin{align}
    \sigma(gg \to h)\times \br(h \to \tau \bar\tau) = 3.1\pm0.2~\pb.
    \label{eq:SMtata}
\end{align}
Note that there is no excess in the $b$-tagging category, 
which implies that the $b$-associated $\phi$ production is disfavored.

Interestingly, 
in the same mass region,
a different excess has also been reported in the di-photon final states by the CMS collaboration 
based on the full Run 1 \cite{CMS:2015ocq} and the first Run 2 (35.9\,fb$^{-1}$) data \cite{CMS:2018cyk}.
The local and global significance are  $2.8\sigma$ and $1.3\sigma$
at $m_{\gamma \gamma} =  95.3\GeV$.
This excess can be interpreted as a new resonance, $\phi$, that decays to two photons, \ie, $gg \to \phi\to\gamma \gamma$ (see Eq.~\eqref{eq:CMS_diphoton}).
Although the similar analysis has been performed by the ATLAS collaboration \cite{ATLAS:2018xad},
the sensitivity is not good enough to check the consistency \cite{Heinemeyer:2018wzl}. 
Moreover, another mild excess has been reported
in the LEP experiment,
that can be interpreted as $e^+e^+\to Z \phi \to Z b\bar{b}$ \cite{LEPWorkingGroupforHiggsbosonsearches:2003ing}.
The signal corresponds to $2.3\sigma$ local significance at $m_\phi=98\GeV$.
Therefore, it is very interesting to consider the possibility that these excesses are caused by the same new particle $\phi$.
 A variety of new physics interpretations have been discussed
in Refs.~\cite{Fox:2017uwr,Haisch:2017gql,Biekotter:2017xmf,Liu:2018xsw,Domingo:2018uim,Hollik:2018yek,Biekotter:2019kde,Cline:2019okt,Cao:2019ofo,Aguilar-Saavedra:2020wrj,Biekotter:2020cjs,Biekotter:2021ovi,Biekotter:2021qbc,Heinemeyer:2021msz,Benbrik:2022azi,Benbrik:2022dja} focusing on the CMS di-photon and the LEP $b\bar{b}$ excesses, while in Refs.~\cite{Biekotter:2022jyr,Biekotter:2022abc} including the CMS di-tau excess as well.

\begin{figure}[t]
\begin{center}
\includegraphics[width=24.5em]{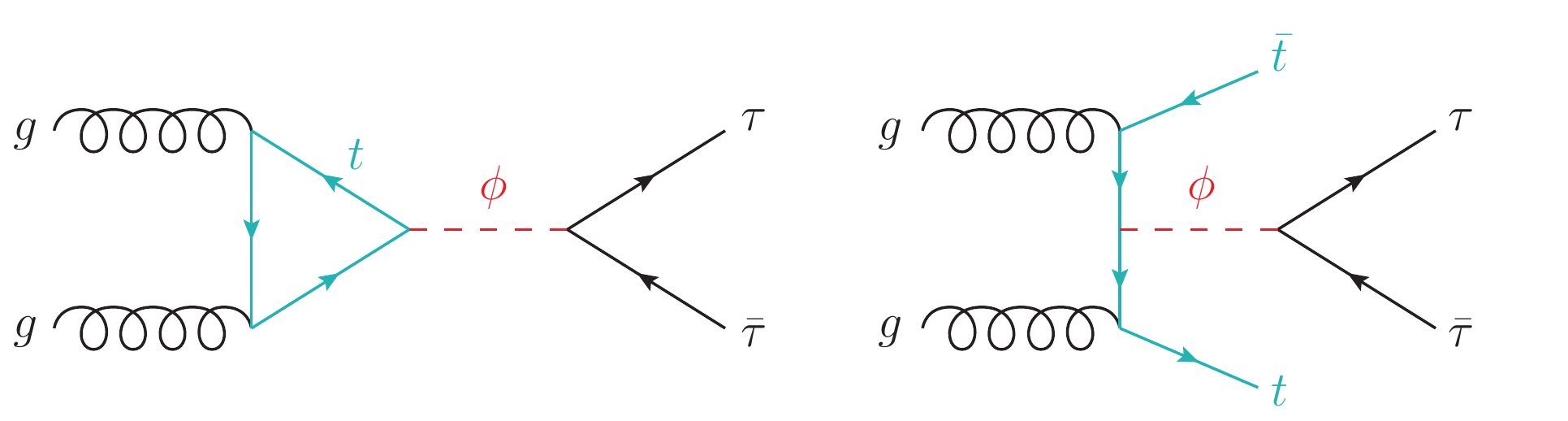}
\includegraphics[width=12.25em]{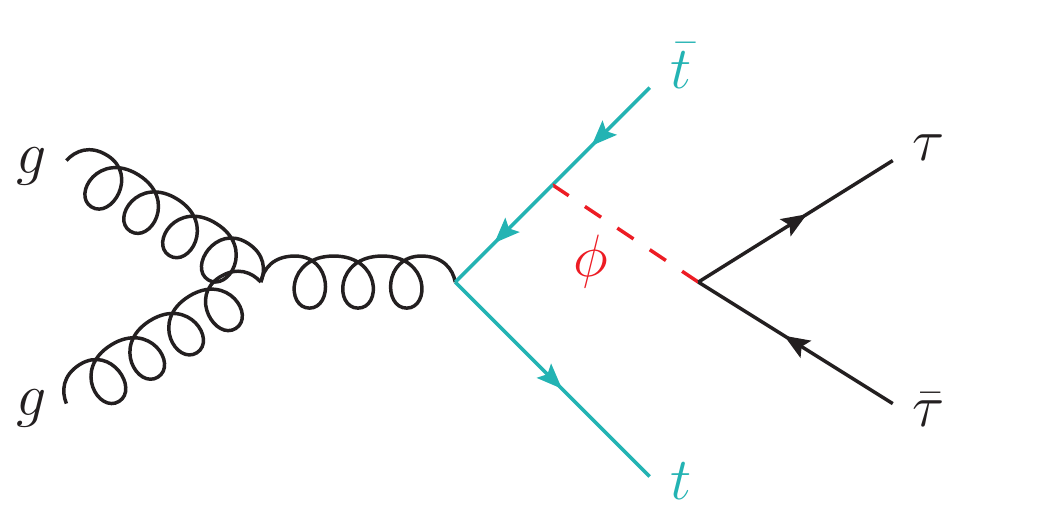}
\caption{
\label{Fig:dia}
Representative Feynman diagrams for a narrow resonance $\phi$ exchange in 
the gluon fusion ($gg \to \tau \bar \tau$) and
the top-quark associated processes ($gg \to t \bar{t} + \tau \bar \tau$).
}
\end{center}
\end{figure}
In this paper,
we point out that associated production of $\phi$ with two top quarks
can provide a simple and powerful way to verify the di-tau excess in Eqs.~\eqref{eq:tata95} and \eqref{eq:tata100}.
In Fig.~\ref{Fig:dia}, 
the relevant diagrams for $gg \to \tau\bar\tau$ 
and $gg\to t\bar{t}+\tau\bar{\tau}$ are shown.
As shown Eq.~\eqref{eq:SMtata}, 
the di-tau excess requires 
the new resonance cross section almost comparable to the SM Higgs boson via the gluon fusion.
The $\tau\bar{\tau}$ resonance at $95$--$100\GeV$, however,
suffers from the huge $Z$ boson background which stems from the tree-level Drell-Yan (DY) process.
Such a $Z$ peak pollution will be mild when the top-quark associated production is selected.
This is because that 
the production cross section of $t\bar{t}+Z$ and $t\bar{t}+h$ are the same size in the SM (see Sec.~\ref{sec:CPeven} for the explicit numbers), and furthermore, the new resonance is produced by the tree level,
which could amplify the sensitivity to probe the new resonance.
We will investigate the LHC sensitivity to the possible $95$--$100\GeV$ resonance in the top-associated process.

In Sec.\,\ref{Sec:setup}, we introduce the minimal setup to account for the $\tau\bar{\tau}$ excess and summarize the predictions of the top-associated productions.
In Sec.\,\ref{Sec:numerics}, the numerical analysis to derive the upper limits on the model-independent $t\bar{t}+\tau\bar{\tau}$ cross section is performed for the new resonance.
Section\,\ref{sec:Summary} is devoted to summary and discussion.

\section{Setup}
\label{Sec:setup}
In this section,
we introduce a simplified model that can account for the excesses in the 95--$100\GeV$ region.
We introduce an additional scalar boson which has large couplings with top and $\tau$, since the gluon fusion is dominated by the chiral heavy fermion loop \cite{Kumar:2012ww} and the resonance preferentially decays into $\tau$ leptons.
It is noted that no excess is found with $\tau\bar\tau$ resonant events in association with $b$ quark jets \cite{CMS:2022rbd}, so that a coupling with $b$ is unlikely to be large.
There are several ways to derive such a scalar effectively at low energy.
In this study, we do not specify models and simply assume that
extra neutral scalars, $H$ and $A$, couple to top and $\tau$ as follows:
\begin{align}
-\mathcal{L}_{\rm eff} =
\frac{\rho_{tt}^H}{\sqrt{2}}\,\bar{t} H t +\frac{\rho_{\tau\tau}^H}{\sqrt{2}}\,\bar{\tau} H \tau
\pm i \frac{\rho_{tt}^A}{\sqrt{2}}\,\bar{t} A \gamma_5 t+i \frac{\rho_{\tau\tau}^A}{\sqrt{2}}\,\bar{\tau} A \gamma_5 \tau\,.
\label{Eq:Lint}
\end{align}
The all couplings, $\rho^\phi_{tt}$ and $\rho^\phi_{\tau \tau}$ ($\phi=H, \, A$), 
are real in our study.
Then, $H$ $(A)$ corresponds to a CP-even (CP-odd) scalar.\footnote{Although additional vector particle,
that couples to the top and $\tau$ \cite{Abdullah:2019dpu}, could also be considered as the candidate, 
the gluon fusion (via the top loop) vanishes for the vector boson production when the fermion interactions are vector couplings \cite{Campbell:2011cu}. We will discuss the possibility of the vector particle interpretation  in Sec.~\ref{sec:Summary}.
}
As we will see, 
the $\gamma_5$ structure plays an important role in our analysis.  
Note that the relative sign for the $A$ interaction depends on the UV theory.
In our study, this relative sign does not affect the conclusion.

The possible UV completion of the light scalar boson
could be the Two-Higgs-Doublet Model (2HDM) with a real singlet scalar~\cite{Biekotter:2019kde,Biekotter:2020cjs,Biekotter:2021qbc,Heinemeyer:2021msz,Biekotter:2022jyr,Biekotter:2022abc},
flavor-aligned 2HDM~\cite{Pich:2009sp}, Generic 2HDM (G2HDM, type-III 2HDM)~\cite{Branco:2011iw,Cline:2015lqp}, axion-like particle (ALP)~\cite{Georgi:1986df}, NMSSM~\cite{Ellwanger:2009dp} and so on.\footnote{As a specific model example,
one can consider the flavor-aligned 2HDM which can approximately reproduce the Lagrangian in Eq.\,(\ref{Eq:Lint}) with $\xi_d=0$, $\xi_\ell=\mathcal{O}(0.1)$ and $\xi_u=\mathcal{O}(0.1)$.
}
If one takes a G2HDM as an illustrative model,
the oblique corrections and the LHC constraint on a charged scalar \cite{ATLAS:2018gfm,ATLAS:2021upq}
do not allow to set both $H$ and $A$ masses around $100\GeV$ for the $\tau\bar{\tau}$ excess. 
In this paper we do not discuss the detailed setup and consider the two cases
that either $H$ or $A$ resides around 100 GeV.
We comment on the other couplings except for the effective Lagrangian in Eq.\,\eqref{Eq:Lint}, which exist in the UV completed models in general.
We suppose only that the gluon-fusion of $\phi$ is dominated in the top-quark loop contribution, which would be correct when the coupling to the bottom quark is not comparable to the top one. Then, a correlation between two cross sections, $\sigma(pp \to gg \to \phi \to \tau \bar{\tau})$ and $\sigma(pp \to t\bar{t} +\phi \to t\bar{t}+\tau\bar\tau)$, becomes robust and is independent of the $\phi$'s decay modes with the other interactions not presented in Eq.~\eqref{Eq:Lint} (including dark sectors).

\subsection{CP-even scalar scenario}
\label{sec:CPeven}
First, we consider the case that the resonance reported by the CMS collaboration is a CP-even scalar $H$. 
The gluon fusion production cross section of $H$ at $\sqrt{s}=13\TeV$ is predicted as follows
\begin{align}
 &\sigma(pp\to gg\to H)= 87.2 \,(\rho_{tt}^H)^2~\pb\quad {\rm{for}}~ m_H=95\GeV\,,
 \label{Eq:ggH95}\\
 &\sigma(pp\to gg\to H)= 79.5 \, (\rho_{tt}^H)^2~\pb\quad {\rm{for}}~ m_H=100\GeV\,,
\label{Eq:coef_ggF_H}
\end{align}
which are evaluated by \texttt{SusHi\,v1.7.0}\,\cite{Harlander:2012pb,Harlander:2016hcx} at next-to-next-to leading order (NNLO).\footnote{%
It is noted that the next-to-NNLO (N$^3$LO) correction increases the cross section up to $3\%$.}
The di-tau excess in Eqs.~\eqref{eq:tata95} and \eqref{eq:tata100} can be accommodated by
\begin{align}
    &\rho_{tt}^H \sqrt{\br(H \to \tau \bar\tau)} = 0.30\pm0.07
    \quad {\rm{for}}~ m_H=95\GeV\,,\\
    &\rho_{tt}^H \sqrt{\br(H \to \tau \bar\tau)} = 0.27\pm0.05
    \quad {\rm{for}}~ m_H=100\GeV\,.
\end{align}
We find that $\rho^H_{tt}$ is sizable to explain the excess: $\rho^H_{tt} \gtrsim 0.22$.

The $\rho_{tt}^H$ interaction also contributes to
the production of $\phi$ in association with two top quarks, as shown in the middle and right diagrams in Fig.~\ref{Fig:dia}.
The production cross section at $\sqrt{s}=13\TeV$ 
is obtained as 
\begin{align}
&\sigma(pp\to t\bar{t}+ H)= 1.07\,(\rho_{tt}^H)^2~\pb\quad {\rm{for}}~ m_H=95\GeV\,,
\label{Eq:ttH95}\\
&\sigma(pp\to t\bar{t}+ H)= 0.94\,(\rho_{tt}^H)^2~\pb\quad {\rm{for}}~ m_H=100\GeV\,,
\label{Eq:prediction_ttH}
\end{align}
at next-to leading order (NLO).
Here, we evaluate the production cross section at the leading order by \texttt{MadGraph5\_aMC@NLO} \cite{Alwall:2014hca} 
and multiply the NLO $K$ factor of 1.29 for simplicity \cite{Frixione:2014qaa}.

Combining the above results, the explanation of the excess predicts
a sizable cross section of $pp \to t\bar{t} H \to t\bar{t} + \tau\bar{\tau}$:
\begin{align}
&\sigma(pp\to t\bar{t}+ H)\times \br(H \to \tau \bar\tau) =[0.056,0.094,0.14]~\pb \quad{\rm{for}}~ m_H=95\GeV\,,
\label{Eq:prediction_H95}
\\
&\sigma(pp\to t\bar{t}+ H)\times \br(H \to \tau \bar\tau) =[0.045,0.069,0.097]~\pb\quad {\rm{for}}~ m_H=100\GeV\,,
\label{Eq:prediction_H}
\end{align}
where numbers in parentheses indicate the $1\sigma$ range  with its central value.

The top-associated production cross section of the SM Higgs boson
has been measured with 79.8 fb$^{-1}$ of the Run 2 data \cite{ATLAS:2018mme}:
 $\sigma (pp\to t\bar{t} + h)=0.67\pm0.14\,\pb$. 
This is consistent with the SM prediction, $\sigma (pp\to t\bar{t}+h)_{\rm SM}=0.51\,\pb$.
Combining the subsequent decay branching ratio of  $\br(h\to\tau\bar\tau)_{\rm SM}\sim6\%$,
$\sigma(pp\to t\bar{t} + h)_{\rm SM} \times\br({h\to\tau\bar\tau})_{\rm SM} \simeq  0.03\,\pb$
is derived in the SM, and thus the predicted cross section of $H$ is larger than the SM Higgs by a factor of approximately three.

We comment on
the top-associated $Z$ production cross section.
It has been measured based on the full Run 2 data \cite{ATLAS:2021fzm}; 
$\sigma (pp\to t\bar{t}+Z)=0.99\pm0.09\,\pb$.\footnote{%
The cross section of $\sigma (pp\to t\bar{t}+Z)$ has been measured in $Z\to \ell\bar{\ell}$ with $\ell=e$ and $\mu$ channels.
}
This result is also consistent with the SM prediction, $\sigma (pp\to t\bar{t}+Z)_{\rm SM} = 0.84^{+0.09}_{-0.10}\,\pb$
at NLO QCD and EW accuracy \cite{Frixione:2015zaa}.
Since $\br(Z\to \ell\bar{\ell})$ is about $3\%$, 
we obtain $\sigma (pp\to t\bar{t}+Z)\times \br(Z\to \tau\bar\tau) \simeq 0.03\,\pb$, 
that is comparable to the top-associated production of $h$.

\subsection{CP-odd scalar scenario}
Next, we consider the CP-odd scalar $A$ interpretation.
The analysis in Sec.~\ref{sec:CPeven} is applied to this case, replacing $H$ with $A$. The sizes of the cross sections are, however, different because of the couplings in Eq.~(\ref{Eq:Lint}), so that the predictions are totally different.

The gluon fusion cross section of $A$ at $\sqrt{s}=13\TeV$ is
\begin{align}
 &\sigma(pp\to gg\to A)= 201.7\,(\rho_{tt}^A)^2\,\pb\quad {\rm{for}}~ m_A=95\GeV\,,\\
 &\sigma(pp\to gg\to A)= 184.4\,(\rho_{tt}^A)^2\,\pb\quad{\rm{for}}~ m_A=100\GeV\,,
\label{Eq:ggF_A}
\end{align}
at NNLO \cite{Harlander:2012pb,Harlander:2016hcx}.
Assuming $\rho^H_{tt}=\rho^A_{tt}$, these predictions are twice as large as them in the $H$ case
due to the different $\gamma_5$ structure in the top-quark loop.
The di-tau excess in Eqs.~\eqref{eq:tata95} and \eqref{eq:tata100} can be accommodated by
\begin{align}
    &\rho_{tt}^A \sqrt{\br(A \to \tau \bar\tau)} = 0.20\pm0.04
    \quad {\rm{for}}~ m_A=95\GeV\,,\\
    &\rho_{tt}^A \sqrt{\br(A \to \tau \bar\tau)} = 0.18\pm0.03
    \quad {\rm{for}}~ m_A=100\GeV\,.
\end{align}

The production cross section of the top-associated production is
\begin{align}
&\sigma(pp\to t \bar{t}+ A)=0.30\,(\rho_{tt}^A)^2\,\pb\quad{\rm{for}}~ m_A=95\GeV\,,\\
&\sigma(pp\to t\bar{t}+ A)=0.29\,(\rho_{tt}^A)^2\,\pb\quad{\rm{for}}~ m_A=100\GeV\,,
\end{align}
where the same NLO $K$ factor as in the CP-even scalar production is assumed \cite{Frederix:2011zi}.
In Ref.~\cite{Frederix:2011zi}, it has been shown that the NLO $K$-factors are almost independent of the CP-even/odd and the mass.
In contrast to the gluon fusion cross section,
the results are three times as small as them in the $H$ case when $\rho^H_{tt}=\rho^A_{tt}$, as shown in Eqs.~\eqref{Eq:ttH95} and \eqref{Eq:prediction_ttH}.
This is again originated from the $\gamma_5$ structure. There is a destructive interference between the contribution of the middle diagram and that of the right diagram in Fig.~\ref{Fig:dia} 
\cite{Djouadi:2005gj,Frederix:2011zi,Dolan:2016qvg}. 

As a result, the cross section of 
$pp\to t\bar{t}+A \to t\bar{t} + \tau\bar{\tau}$, that is consistent with the di-tau excess, is predicted to be smaller than that of the $H$ case:
\begin{align}
&\sigma(pp\to t\bar{t}+ A) \times \br(A \to \tau \bar\tau)=[0.007,0.011,0.017]\,\pb \quad{\rm{for}}~ m_A=95\GeV\,,\label{Eq:ttA95}\\
&\sigma(pp\to t\bar{t}+ A)\times \br(A \to \tau \bar\tau)=[0.005,0.009,0.013]\,\pb \quad{\rm{for}}~ m_A=100\GeV\,.
\label{Eq:prediction_A}
\end{align}
It is also known that the angular correlation and transverse momentum distributions of $pp\to t\bar{t}+A$ are different from $pp\to t\bar{t}+H$ due to the presence of $\gamma_5$ in Eq. (\ref{Eq:Lint}) \cite{Dolan:2016qvg}.

\section{Comparisons to the ATLAS data}
\label{Sec:numerics}

In this section, 
we compare the CMS di-tau excesses in Eqs.~\eqref{eq:tata95} and \eqref{eq:tata100} to the ATLAS results that involve $\tau\bar{\tau}$ in the final states. 
Since the range of the ATLAS data set for the exotic particle search decaying to $\tau\bar{\tau}$ is $m_{\tau\tau}\geq 200\GeV$  \cite{ATLAS:2020zms}, the $95$--$100\GeV$ region has not been covered.
Instead, we utilize the ATLAS Run 2 full data for the $h \to \tau\bar\tau$ decay channel where $h$ is the SM Higgs boson \cite{ATLAS:2022yrq}.
The $h\to\tau\bar{\tau}$ and $Z\to\tau\bar{\tau}$ events have been carefully studied by using the several production processes.
Therefore the additional narrow resonance $\phi$ that decays to $\tau\bar{\tau}$ can also be probed.
In the following, we analyze the data in the {\em boosted $\tau_{\rm h} \tau_{\rm h}$} and {\em $tt(0\ell) + \tau_{\rm h} \tau_{\rm h}$ categories}~\cite{ATLAS:2022yrq}, where $\tau_{\rm h}$ denotes $\tau$ that decays hadronically.

\subsection{Boosted \texorpdfstring{$\tau\bar{\tau}$}{tau tau} search}
\label{sec:boostedtautau}

The boost categories defined in Ref. \cite{ATLAS:2022yrq} consist of the events that fail to meet the criteria of the vector-boson fusion (VBF), vector-boson associated production (VH), and a pair of top-quark associated production. The events have high-$p_{\rm T}$ (boosted) Higgs candidates. 
The more than $70 \%$ events, actually, come from the gluon fusion with large Higgs boson transverse momentum in the analysis of Ref.~\cite{ATLAS:2022yrq}. 
Hence, the ATLAS data in this category should be a good comparison to the CMS di-tau excess.
We use the result in the {\em boost\_2~category}~corresponding to  $200 < p_{\rm T}^{\tau\tau} < 300\GeV$ in Ref.~\cite{ATLAS:2022yrq}.\footnote{%
We found that the data in {\em boost\_3~category}~corresponding to  $ p_{\rm T}^{\tau\tau} > 300\GeV$ is less sensitive in the following analysis due to the small amount of statistics.}

\begin{figure}[t]
\begin{center}
\includegraphics[width=17em]{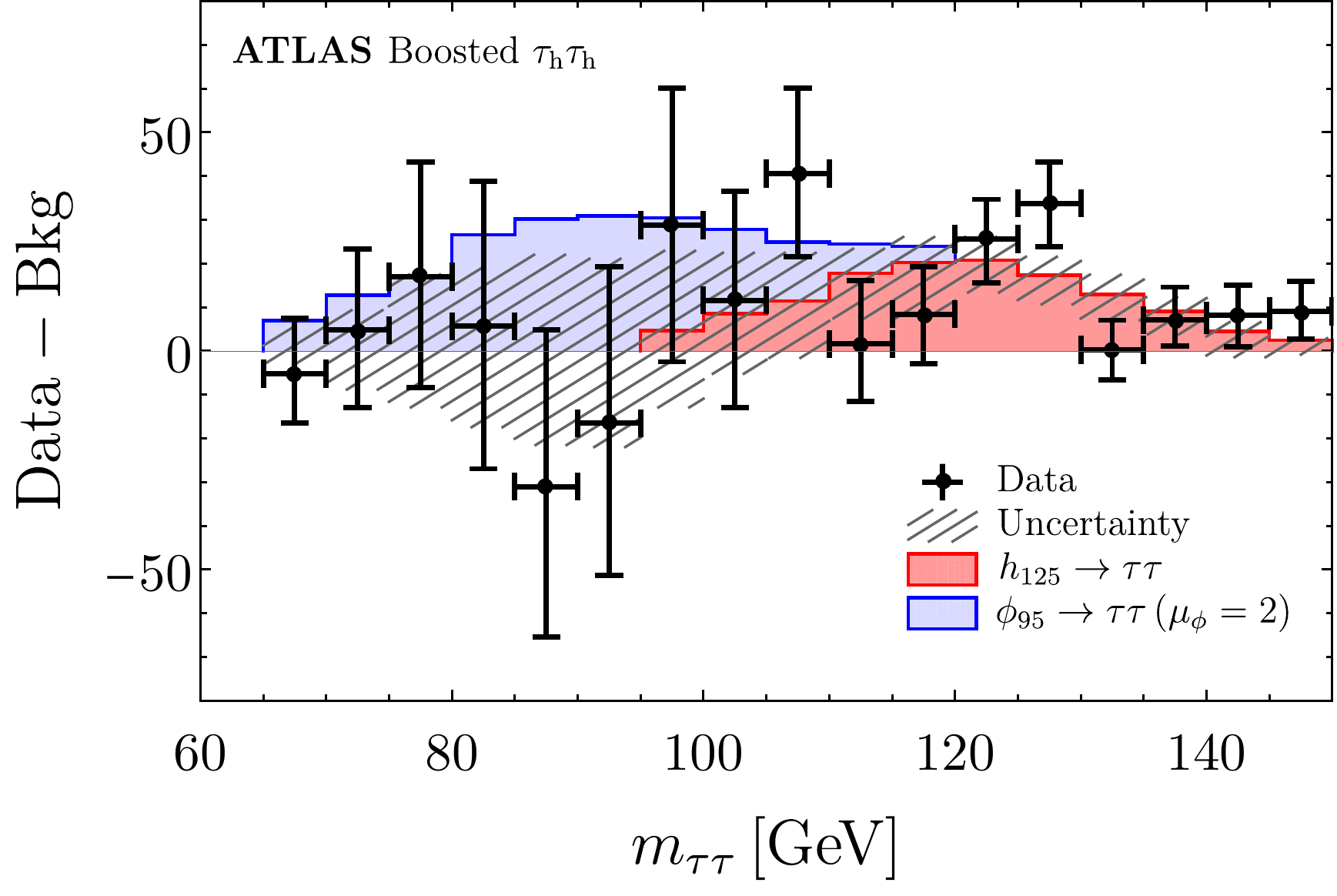}
\quad
\includegraphics[width=17em]{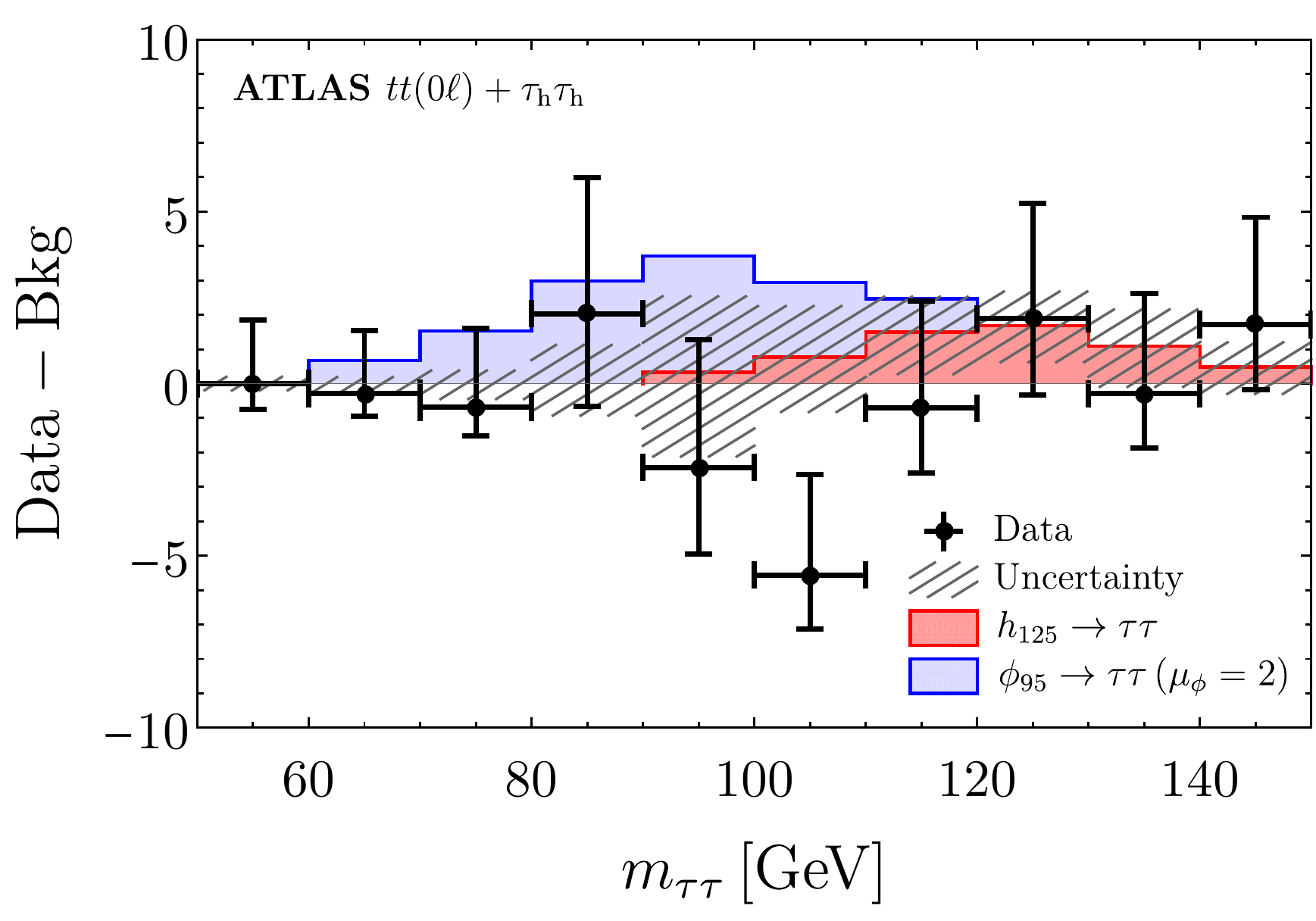}
\caption{
\label{Fig:95_mu2}
The histograms of the additional resonance $\phi$ (blue)
and SM Higgs (red) event shapes are shown with the experimental data (cross) and their uncertainties (dashed band).
The left panel is for the boosted $\tau_{h}\tau_{h}$ and the right panel is
the $tt(0\ell) + \tau_{\rm h}\tau_{\rm h}$ categories, respectively.
The signal normalization is assumed to be $\mu_\phi=2$, and  $m_\phi=95\,$GeV are set for the mass.
See the text for details.
}
\end{center}
\end{figure}

Using the the narrow width approximation, 
we define a signal strength for the gluon fusion of $\phi$,
\begin{align}
\mu_\phi (\tau\bar\tau) \equiv 
\frac{\sigma(pp\to gg \to \phi)\times \br(\phi\to \tau\bar\tau)}{\sigma(pp\to gg \to h)_{\rm{SM}}\times \br(h\to \tau\bar\tau)_{\rm{SM}}}\,,
\end{align}
where the denominator is values of the $125\GeV$ SM Higgs.
In the left panel of Fig.~\ref{Fig:95_mu2}, an expected histogram of the $\phi$ with $m_{\phi}=95\GeV$ and $\mu_{\phi}(\tau\bar{\tau})=2$ is shown by the blue shaded region.
Here the SM Background (Bkg), except for the SM Higgs boson, is subtracted from the data which corresponds to 
the bottom-left panel of Fig.~20 ({\em boost\_2~category}) of Ref.~\cite{ATLAS:2022yrq}.
Note that the SM Higgs histogram (red shaded region) stands for $pp \to h \to \tau \bar{\tau}$ and is scaled by $1/0.93$ from the original figure.\footnote{%
We also subtract the non gluon-fusion contributions according as Table~11 of Ref.~\cite{ATLAS:2022yrq} to construct the histogram of $\phi$. 
} 
The $0.93\,(^{+0.13}_{-0.12})$ is a global fit result of the signal strength of $p p \to h \to \tau \bar{\tau}$ \cite{ATLAS:2022yrq}.
The dashed band represents the total uncertainty of the SM Bkg. 
It is shown that there is a huge uncertainty around $m_{\tau\tau}=80$--$100\GeV$, which comes from the DY (boosted) $Z$-boson production.
%

By using a $\chi^2$ test which will be explicitly defined in the next section,
we set $95\%$ confidence level (CL) upper limits on the signal strength,
\begin{align}
    & \mu_\phi (\tau\bar\tau) < 3.78 \quad \text{for~}m_{\phi} = 95\GeV\,,
    \\
        & \mu_\phi (\tau\bar\tau) < 3.61  \quad \text{for~}m_{\phi} = 100\GeV\,,
\end{align}
or equivalently, 
the upper limits on the production cross sections, 
\begin{align}
&\sigma(gg \to \phi)\times \br(\phi \to \tau\bar\tau) < 11.5
\,\pb\quad \text{for~}m_{\phi} = 95\GeV\,,
\\
&\sigma(gg \to \phi)\times \br(\phi \to \tau\bar\tau) < 11.0
\,\pb\quad \text{for~}m_{\phi} = 100\GeV\,.
\end{align} 
We find that the ATLAS data in the boosted $\tau_{\rm h}{\tau}_{\rm h}$ category 
are consistent with the CMS di-tau excess in Eqs.~\eqref{eq:tata95} and \eqref{eq:tata100} even if the upper edge of the 1$\sigma$ is considered.

\subsection{\texorpdfstring{$t\bar{t}+\tau\bar{\tau}$}{t t + tau tau} search}
\label{sec:tttautau}

In this section, we compare the CMS di-tau excess with the ATLAS data in the $tt(0\ell) + \tau_{\rm h} \tau_{\rm h}$ category \cite{ATLAS:2022yrq}. 
Due to the limited statistics,
the ATLAS collaboration has not performed a serious top reconstruction,
but just imposed either six jets including at least one $b$-tagged jet or five jets including at least two $b$-tagged jets
for the event selection.

As mentioned in Sec.~\ref{sec:intro}, 
the additional $t\bar{t}$ requirement is expected to improve the sensitivity to probe the new resonance, because the huge DY $Z$-boson Bkg can be vetoed and the production cross sections of $t\bar{t} + \{h,Z,\phi\} \to t\bar{t} + \tau\bar{\tau}$ are of the same size. 
However, since the top-quark tagging has  not been seriously imposed in this ATLAS data, 
the SM Bkg is still dominated by the DY process ($pp \to Z+5\text{--}6\,$QCD~jets), 
and still large Bkg  remains even in the signal region ($m_{\tau\tau}\simeq 125\GeV$),
as one can see in Fig.~11 of Ref.~\cite{ATLAS:2022yrq}.
The region of interest in this paper is 
 $m_{\tau\tau}=95$--$100\GeV$,
 and hence we have the large Bkg from $Z\to \tau\bar{\tau}$.
Nonetheless, 
we use this result to derive the current experimental limit on the new resonance production.
It is naively expected that a severe top-quark tagging algorithm (by the  mass reconstruction)
 improves the sensitivity.


Since experimental analyses have used the boosted decision tree (BDT) techniques, 
it is difficult to access the detailed information about the final kinematic cuts.
Instead, we utilize the data in {\em ttH\_1~category}, corresponding to the right panel of Fig.~11 of Ref.~\cite{ATLAS:2022yrq}, where the $t\bar{t}h$ events are optimised to be enhanced over $Z$ and $t\bar{t}$ Bkg events by the BDT.
In the plot,
$(\text{data~yields)}-(\text{SM~Bkg~except~for~the~SM~Higgs})$ is shown ($N_{\rm ob} \pm \Delta N_{\rm ob}$), as well as 
uncertainty of the total SM Bkg ($\pm \Delta N_{\rm Bkg}$) and the SM Higgs histogram ($N_{h}$).
From this figure, we estimate the sensitivity to probe the new resonance,
supposing that the $m_{\tau\tau}$ distribution of  $t\bar{t} +\phi \to t\bar{t} + \tau\bar{\tau}$  under the BDT is similar to the SM Higgs.

Similar to the previous section, 
we define a signal strength for the top-associated production of $\phi$,
\begin{align}
\mu_\phi (t\bar{t}+\tau\bar\tau) \equiv 
\frac{\sigma(pp\to t\bar{t}+\phi)\times \br(\phi\to \tau\bar\tau)}{\sigma(pp\to t\bar{t}+h)_{\rm{SM}}\times \br(h\to \tau\bar\tau)_{\rm{SM}}}\,.
\end{align}
In the right panel of Fig.~\ref{Fig:95_mu2},
an expected histogram of $\phi$ with $m_{\phi}=95\GeV$ and $\mu_{\phi}(t\bar{t}+\tau\bar\tau) =2$ 
is shown by the blue shaded region. 
Since the each bin width in the data is $10\GeV$, 
we simply shift the distribution of the SM Higgs histogram
by three bins to obtain the distribution of $m_\phi=95\GeV$.\footnote{%
See also Sec.~\ref{sec:Summary} for a discussion of this treatment.}
Since the width of the SM Higgs histogram stems from the experimental resolution, 
it is expected that the width of the histogram of $\phi$ is roughly of the same size as the SM Higgs. 
Moreover, if the resolution is proportional to the value of $m_{\tau \tau}$, the histogram of $\phi$ becomes sharpened.
Therefore, we just rescale the SM Higgs histogram to conservatively predict  the $\phi$ contribution. 
We represent the histogram of $\phi$ by $N_\phi(\mu_\phi)$.
Again, the SM Higgs histogram (red shaded region)
is scaled by $1/0.93$ from the original one.

We perform the following $\chi^2$ test,
\begin{align}
\chi^2 (\mu_\phi) = \text{max} \left[\chi^2_i(\mu_\phi) \right] \quad \text{and} \quad
    \chi^2_i(\mu_\phi)=\sum_{j=i}^{i+2} \frac{\left[N_{\rm ob}^j -N_{h}^{j} - N_{\phi}(\mu_\phi)^j \right]^2}{(\Delta N_{\rm ob}^{j})^2+(\Delta N_{\rm Bkg}^j)^2} 
    \,,
    \label{eq:chisq}
\end{align}
where $i$ and $j$ are indices of each bin.
Note that since correlations between the data in each bin are not available in Ref.~\cite{ATLAS:2022yrq}, we discard them for simplicity.
Due to the finite experimental resolution for the $m_{\tau\tau}$ distribution, 
judging based on the single bin data is too aggressive.
Here, we use at least $3$ contiguous bins for each $\chi^2_i(\mu_\phi)$ evaluation.\footnote{%
We found that $\chi^2$ tests for $\mu_{\phi}(t \bar{t}+\tau\bar\tau)$ with 2 contiguous bins give roughly 30$\%$  stronger constraints
for both $m_\phi=$ 95\,\GeV and 105\,GeV,
while the analyses using 4 contiguous bins bring 25$\%$ and 40$\%$ weaker constraints for 95\,\GeV and 105\,\GeV, respectively.}
The criterion for setting the upper limit on $\mu_\phi$
is $\chi^2(\mu_\phi) < \chi^2_{3\text{dof}, 95\%} \simeq 7.82$.

As a validation of this $\chi^2$ test,
we compare the upper limit on the SM Higgs production cross section in the ttH\_1~category.
We obtain the 95\% CL upper limit for the Higgs boson cross section,
\begin{align}
  \mu_h =  \frac{\sigma(pp\to t\bar{t}+h) \times \br (h \to \tau\bar{\tau})}
    {\sigma(pp\to t\bar{t}+h)_{\rm SM} \times \br( h \to \tau\bar{\tau})_{\rm SM}}\le 2.65\,,
\end{align}
while $\mu_h \lesssim 2.96$ ($\mu_h=1.02^{+0.97}_{-0.81}$) has been set in Ref.~\cite{ATLAS:2022yrq}.
It is found that the $\chi^2$ test in Eq.~\eqref{eq:chisq} gives a slightly severe limit.
This could be attributed to the following reasons;
This $\chi^2$ test does not include the theoretical uncertainty of the signal events properly. 
The total uncertainty (dashed band $\Delta N_{\rm Bkg}$) includes the theoretical uncertainty of $\mu_h \simeq 1$ and it should be inflated according as $\mu_h$.
Second, when we evaluate the $\chi^2_i$ value for at least not $3$ but $4$ contiguous bins, the resultant limit is weakened by $5$--$10\%$.
Therefore, we decide to weaken the obtained upper limit on $\mu_{\phi}$
 by
$10\%$ 
to be conservative test.
In this prescription, we obtain $\mu_h < 2.92$.

\begin{table}[t]
    \centering
    \renewcommand{\arraystretch}{1.5}
    \newcommand\TIMES{\!\times\!}
    \newcommand{\bhline}[1]{\noalign{\hrule height #1}}
    \rowcolors{2}{gray!15}{white}
    \begin{tabular}{lccc}
    \bhline{1pt}
    &$m_{\phi}=95\GeV$&$m_{\phi}=100\GeV$&$m_{\phi}=105\GeV$\\
    \hline
    $\mu_{\phi}(\tau \bar{\tau})$ &$<3.78$&$<3.61$& $<2.00$\\
    $\sigma(gg\to\phi)\TIMES\br(\phi\to\tau\bar\tau)$ &$<11.5\,\pb$&$<11.0\,\pb$& $<6.08\,\pb$\\
    $\mu_{\phi}(t\bar{t}+\tau\bar{\tau})$ &$<1.56$&--& $<1.10$\\
    $\sigma(pp\to t\bar{t} +\phi)\TIMES\br(\phi\to\tau\bar\tau)$ &$<0.050\,\pb$&--& $<0.035\,\pb$\\
    \bhline{1pt}
    \end{tabular}
    \caption{\label{Tab:result_muphi}
    The $95\%$ CL upper limits on the signal strengths ($\mu_\phi$) and the production  cross sections for $m_{\phi}=95,\,100,\,105\GeV$ are summarized. 
    These limits are obtained from the ATLAS Run 2 full data \cite{ATLAS:2022yrq} in the boosted $\tau\bar{\tau}$ (boost\_2~category) and $t\bar{t} +\tau\bar{\tau}$ ($tt(0\ell) + \tau_{\rm h} \tau_{\rm h}$ category) searches,  via the $\chi^2$ test defined in Eq.~\eqref{eq:chisq}. }
\end{table}    

Using the $\chi^2$ test, we set the $95\%$ CL upper limit on the signal strength, 
\begin{align}
\mu_\phi (t\bar{t}+\tau\bar{\tau}) < 1.56 
\quad \text{for~}m_{\phi} = 95\GeV\,,
\label{eq:tttata95}
\end{align}
or equivalently, 
\begin{align}
\sigma(pp \to t\bar{t} + \phi)\times \br(\phi \to \tau\bar\tau) < 0.050
\,\pb\quad \text{for~}m_{\phi} = 95\GeV\,.
\end{align} 
It is found that the CP-even scalar $H$ interpretation on the CMS di-tau is excluded
by the ATLAS $t\bar{t}+\tau\bar{\tau}$ search, while the CP-odd $A$ one is consistent with the ATLAS  
(see Eqs.~\eqref{Eq:prediction_H95} and \eqref{Eq:ttA95}).
Note that
due to the fixed bin width, 
we can not perform the same analysis for the $m_\phi=100\GeV$ case.\footnote{%
We performed the same $\chi^2$ test for $m_\phi = 105\GeV$, and the result is $\mu_{\phi}(t\bar{t} +\tau\bar{\tau}) < 1.10$, which is not far from Eq.~\eqref{eq:tttata95}.
This result implies that the upper limit for $m_\phi = 100\GeV$ would be the same size as the $m_\phi = 95\GeV$ case.}
Our results are summarized in Table~\ref{Tab:result_muphi}.

\begin{figure}[t]
\begin{center}
 \includegraphics[width=17.3em]{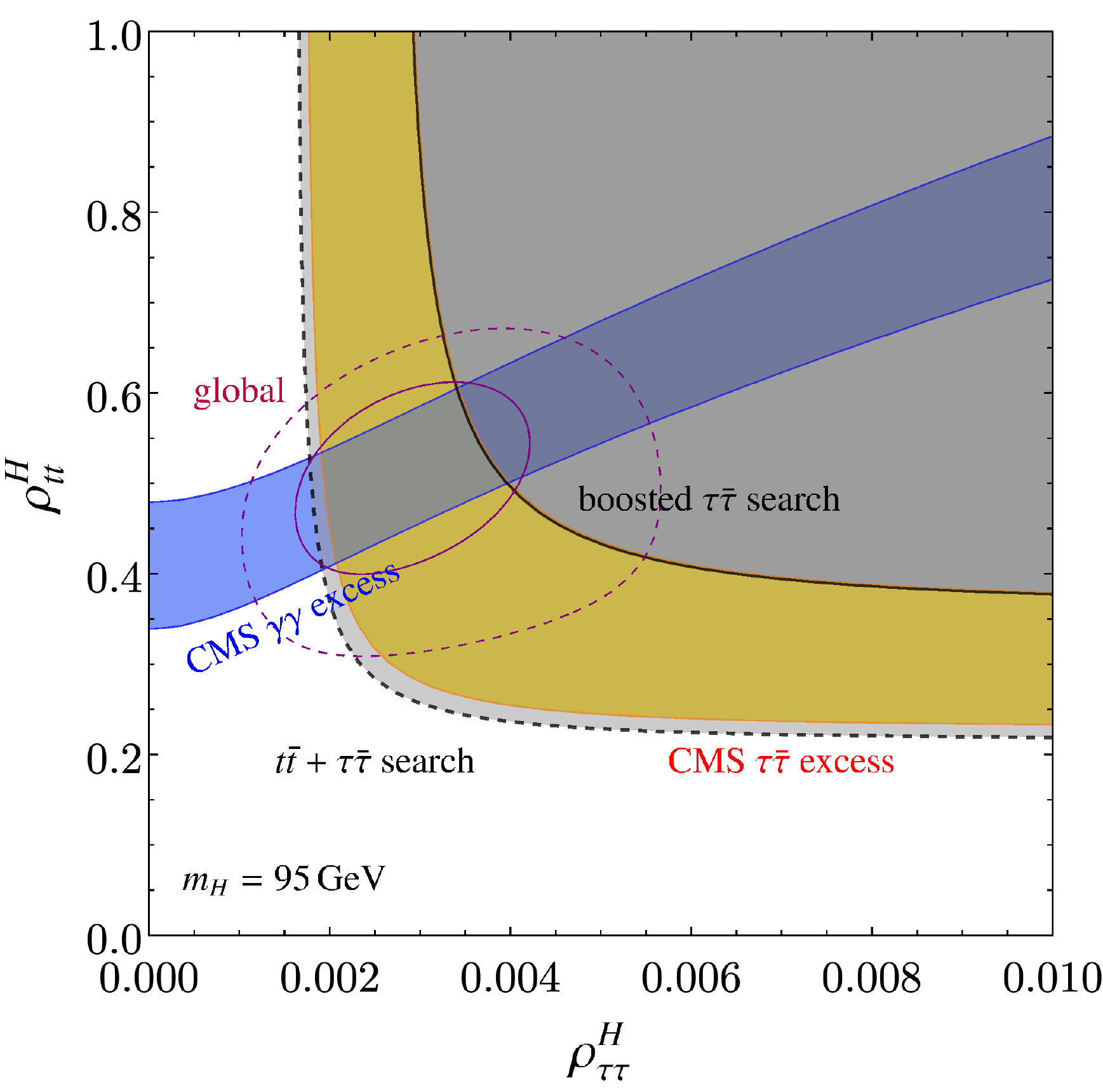}
\quad
\includegraphics[width=17.3em]{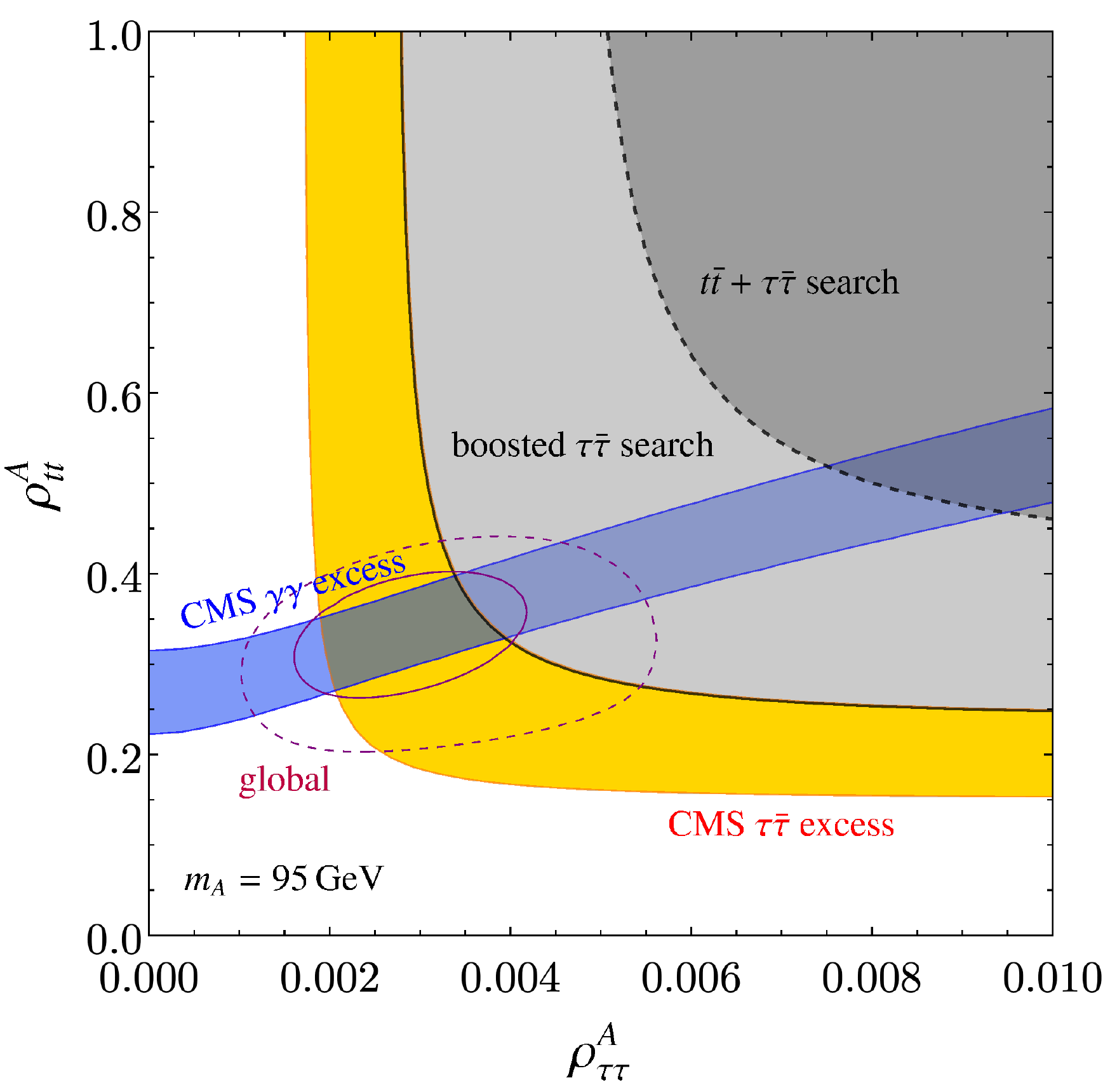}
\caption{
\label{Fig:coupling}
The obtained ATLAS $95\%$ CL limits are shown on a $\rho_{\tau\tau}$--$\rho_{tt}$ plane (see Table~\ref{Tab:result_muphi}). 
The gray shaded regions surrounded by
the solid and dashed lines are excluded
by the ATLAS boosted $\tau\bar{\tau}$ and $t\bar{t}+\tau\bar{\tau}$ searches, respectively.
The left panel is for the CP-even ($H$) case and the right one is for the CP-odd ($A$).
The CMS di-tau and di-photon excesses can be explained at the $1\sigma$ level in the yellow and blue regions, respectively. 
The global fits at the $1\sigma$ ($2\sigma$) level are shown by the purple solid (dashed)  circles.
The scalar boson mass is set to be $95\GeV$.
}
\end{center}
\end{figure}

Finally, we project the obtained ATLAS limits 
onto the the minimal setup in Eq.~\eqref{Eq:Lint}.
In Fig.~\ref{Fig:coupling},
we show the ATLAS limits from the data in the boosted $\tau\bar\tau$ and $t\bar{t}+\tau\bar\tau$ searches 
by the gray shaded regions surrounded by the solid and dashed lines, respectively. 
The CP-even (odd) scalar with $m_{H(A)}=95\GeV$ is considered in the left (right) panel.
The CMS di-tau excess in Eq.~\eqref{eq:tata95} can be explained in the yellow region.
For the branching ratio, 
we calculate $H(A) \to \tau \bar{\tau}$ at the tree level and 
$gg,\,\gamma\gamma$ and $Z\gamma$ at the one-loop level \cite{Gunion:1989we} for the decay channels. 
Furthermore, the aforementioned CMS di-photon excess can be explained in 
the blue region corresponding to \cite{CMS:2018cyk,Biekotter:2019kde}
\begin{align}
    \sigma (gg \to \phi)\times \br(\phi \to \gamma \gamma) = 0.058\pm0.019\,\pb\,,
    \label{eq:CMS_diphoton}
\end{align}
where the \texttt{SusHi} and Ref.~\cite{Denner:2011mq} are used for the $95\GeV$ SM-like Higgs value.
It is clearly shown that an interesting parameter region that can explain both di-tau and di-photon excesses is almost excluded in the CP-even scalar case. 
Note that although the results on Fig.~\ref{Fig:coupling} assume the minimal setup in Eq.~\eqref{Eq:Lint}, the statement that the anomaly-driven  parameter region is excluded in the CP-even scalar case is independent of the effective Lagrangian as long as the gluon-fusion is dominated in the top-quark contribution.
Additional interactions including dark sectors reduce only $\text{BR}(\phi \to \tau\bar\tau)$. This effect shifts the results to the upper-right direction on Fig.~\ref{Fig:coupling}, but the relative positions do not change.

\subsection{Comment on \texorpdfstring{$t\bar{t}+\gamma\gamma$}{t t + gamma gamma} search}
\label{Sec:ttgamgam}
Motivated by the CMS di-photon excess in Eq.~\eqref{eq:CMS_diphoton},
we point out that the above procedure can be repeated
to estimate the upper limit on $\sigma(pp \to t\bar{t}+\phi)\times \br(\phi \to \gamma\gamma)$ 
by analysing $pp \to t\bar{t} + h \to t\bar{t} + \gamma \gamma$ process.
This process must be cleaner than the $pp\to t\bar{t} + \{h,Z\} \to t\bar{t}+\tau\bar{\tau}$.
This is because the $Z \to \gamma\gamma$ decay is forbidden (the Landau--Yang theorem \cite{Landau:1948kw,Yang:1950rg})
so much smaller SM Bkg is expected around $m_{\gamma\gamma} \sim 90\GeV$.\footnote{%
Still, there is a doubly photon-misidentified Bkg from $Z \to e^+e^-$ \cite{CMS:2018cyk}.}
Furthermore, $\sigma(pp \to t\bar{t}+\phi)\times \br(\phi \to \gamma\gamma)$ is expected to be the same size as the 
$\sigma(pp \to t\bar{t}+h)_{\rm SM}\times \br(h \to \gamma\gamma)_{\rm SM}$.
Therefore, the search for the additional resonance in $pp \to t\bar{t} + \gamma\gamma $ is promising.

Indeed, such a check has been implicitly done in the CMS collaboration \cite{CMS:2018cyk}.
However, 
we could not find any results of direct searches for $pp \to t \bar{t} + \gamma \gamma$ in a region of $m_{\gamma\gamma}\le105\GeV$  \cite{ATLAS:2018mme,ATLAS:2020ior,CMS:2021kom} (see, \eg, Fig.~2 of Ref.~\cite{ATLAS:2020ior}).
Therefore, we would like to suggest 
an experimental analysis of the $pp \to t \bar{t} + \gamma \gamma$ process with the low-mass region 
to probe the additional resonance.

\section{Summary and discussion}
\label{sec:Summary}

Due to the nature of the high-energy proton collider 
and its harsh hadron activity,
it is difficult to probe weakly interacting colorless new particles below $\mathcal{O}(100)\GeV$ mass region. 
Very recently, the
CMS collaboration has reported a di-tau excess with a local significance of 2.6--3.1$\sigma$ around $m_{\tau\tau}=95$--$100\GeV$.
This excess can be interpreted as an additional scalar boson $\phi$ produced via the gluon-fusion process.
Interestingly, 
in the same mass region, two other excesses have been reported;
a di-photon excess by the CMS collaboration and 
an excess in $e^+e^- \to Z b\bar{b}$ from the LEP experiment.

In this paper, we focus on the CMS di-tau excess.
 This excess can be explained by the light scalar $\phi$ which couples with the top and $\tau$ in the minimal setup.
 First, we found that the minimal scalar model is still consistent with the ATLAS result for the similar boosted $\tau\bar{\tau}$ search, where a huge SM Bkg comes from the $Z$-boson DY production.
 Second, we point out that the minimal scalar model predicts the inevitable correlation with the top-quark associated process ($gg \to t\bar{t} +\tau\bar\tau$), see the Feynman diagrams in Fig.~\ref{Fig:dia}.
Moreover, a requirement of an additional top-quark pair in the final state suppresses the huge $Z$-boson DY Bkg, so that the experimental sensitivity to probe $\phi$ is certainly better compared to the boosted $\tau\bar{\tau}$ search around $m_{\tau\tau}=95$--$100\GeV$, although the statistics are limited.

Based on the ATLAS data available in Ref.~\cite{ATLAS:2022yrq}, 
we obtain the $95\%$ CL upper limits on the gluon-fusion and  top-associated production cross sections in the minimal setup, which are summarized in Table~\ref{Tab:result_muphi}.
One of important points is that the limits depend on the CP eigenstate of $\phi$, \ie, whether $\phi$ is CP-even ($H$) or CP-odd ($A$).
We point out that 
the gluon-fusion production cross section is twice larger in the $A$ case, while the top-associated production cross section is three times larger in the $H$ case including the QCD higher-order corrections.
This difference is caused by the $\gamma_5$ structure in the Yukawa interaction.
As the result, it is found that
the $H$ interpretation on the di-tau excess is excluded by the ATLAS $t\bar{t}+\tau\bar{\tau}$ search, while the $A$ case is allowed, see Fig.~\ref{Fig:coupling}.
The available ATLAS data (the $tt(0\ell) + \tau_{\rm h} \tau_{\rm h}$ category) is pure hadronic final states.
We also hope the ATLAS collaboration to stop categorizing 
and to combine both leptonic \cite{ATLAS:2017ztq,ATLAS:2019nvo} and hadronic decay modes \cite{ATLAS:2022yrq} 
to increase the sensitivity to probe a possible new particle.
The similar analysis could also be applied in $t\bar{t}+\gamma\gamma$ search to probe a possible light resonance that can accommodate the CMS di-photon excess.

Note that the LEP $b\bar{b}$ excess implies that $\phi$ is the CP-even state ($H$) because the $Z$-boson-associated $A$ production vanishes.
Therefore, once the $b\bar{b}$ excess is involved seriously, 
the CP symmetry must be violated in the scalar model.
Such CP-violating parameters (Yukawa or scalar self interactions) would induce the electron electric dipole moment at two-loop level (\eg, Ref.~\cite{Fuyuto:2015ida}).
Furthermore, a search for $t\bar{t}+b\bar{b}$ signature would be interesting in light of the $b\bar{b}$ excess.
Although Fig.~12 of Ref.~\cite{ATLAS:2017fak} would be helpful to check  the $b\bar{b}$ excess, 
the measurement suffers from the QCD jets.
The more statistics and dedicated study are necessary to make a clear  conclusion.

An additional vector particle ($A'$) would be also a possibility for the CMS di-tau excess. The simplest possibility is the hidden-photon model \cite{Fayet:2007ua,Pospelov:2008zw}. 
The production of $A'$ is the DY process and the decay branching ratio is fixed [$\br(A'\to\tau\bar{\tau})\simeq 15\%$] \cite{Buschmann:2015awa}.
Currently there is no direct experimental bound on the $A'$ production with mass region of around 90\,GeV at the LHC \cite{CMS:2019buh}.
Instead, an indirect bound comes from measurements of the EW precision observables at the LEP and Tevatron experiments 
\cite{Curtin:2014cca}.
It is, however, beyond the scope of this paper to examine whether the $A'$ interpretation is allowed.

We also comment on our prescription for the $\chi^2$ test in Eq.~\eqref{eq:chisq}.
Our prescription is intuitive and does not rely on realistic Monte-Carlo simulations.
Necessary cut information for the detailed analysis
is not available from the experimental papers
because the BDT algorithm is adopted.
Our prescription may receive additional effects from the BDT, 
since the BDT variables include the sub-leading $p_{\rm T}^\tau$ and missing transverse momentum.

Since the current ATLAS data is consistent with the CMS di-tau excess if  a light CP-odd scalar is introduced,
it is nice to consider specific new physics models.
A light CP-odd scalar emerges as a pseudo-Nambu--Goldstone (NG) boson which comes from the spontaneous symmetry breaking of some global symmetries.
Such a mass of the pseudo-NG boson does not lead to an additional fine-tuning problem.

Within the G2HDM, additional scalars appear in the same mass scale with moderate mass differences of $\mathcal{O}$(VEV) (see, Ref.~\cite{Iguro:2017ysu} for instance).
It is possible to predict the lighter CP-odd scalar in general.
Nevertheless the mass degenerated setup, where only $\rho_{tt}$ and $\rho_{\tau\tau}$ are assumed to be nonzero, is not compatible with the direct search for the charged scalar in $pp\to tb + H^\pm\to tb+\tau\nu~{\rm{or}}~tb$ \cite{ATLAS:2018gfm,ATLAS:2021upq}.
Therefore the heavier charged scalar is necessary while keeping the CP-odd scalar mass to be around 95$\GeV$.\footnote{The mass degeneracy of a heavier CP-even scalar and a charged scalar is required to satisfy the constraint from the electroweak precision test.} 
Another possible solution to this dilemma is to put the additional coupling and open up a new decay mode of a charged scalar.
However, additional Yukawa couplings could also contribute to the scalar production and suffers from the direct searches.
The weakest constrained Yukawa coupling is a top-charm flavor violating coupling, $\rho_{tc}$, where top quark is a left handed and charm quark is right handed. 
Since the top mass is heavier than 100\,GeV, the coupling does not reduce the BR($\phi\to\tau\bar{\tau}$). 
On the other hand the $SU(2)_L$ rotation generates $H^-\to b\bar{c}$ decay when $\rho_{tc}$ is not vanishing.
It is recently pointed out that  low-mass di-bottom jets would be sensitive to this coupling depending on the mass \cite{Iguro:2022uzz}.
It is worthwhile to comment that the non-zero product of $\rho_{tc}$ and $\rho_{\tau\tau}$ can enhance  $R_{D^{(*)}}={\rm{BR}}(\bar{B}\to D^{(*)} \tau\bar{\nu})/{\rm{BR}}(\bar{B}\to D^{(*)} \ell\bar{\nu}$), where $\ell=e,\,\mu$ and the 3--4$\sigma$ discrepancy is reported \cite{Aoki:2021kgd,Iguro:2020cpg},
consistently with the $B_c$-meson lifetime \cite{Blanke:2022pjy}.\footnote{For $m_{H^{-}}\ge400$\,GeV, $\tau\nu$ resonance searches exclude the interpretation \cite{Iguro:2018fni}.}
The non-zero $\rho_{tc}$ with the large mass difference between neutral scalars induces the same sign top signal \cite{Iguro:2018qzf}. This is also interesting as well as the search for a light scalar in the double-scalar production at the LHC \cite{ATLAS:2021fet}.
The more quantitative and dedicated study is beyond this paper and will be given elsewhere.

\section*{Acknowledgements}
S.\,I. would like to thank Shigeki Hirose, Yuta Takahashi, Ulrich Nierste, Andreas Crivellin, and Hantian Zhang for fruitful comments and valuable discussion.
T.\,K. thanks 
Junji Hisano, Kazuhiro Tobe, and Nodoka Yamanaka
for useful discussion.
The work of S.\,I.~is supported by the Deutsche Forschungsgemeinschaft (DFG, German Research Foundation) under grant 396021762-TRR\,257.
T.\,K.~is supported by the Grant-in-Aid for Early-Career Scientists (No.\,19K14706) from the Ministry of Education, Culture, Sports, Science, and Technology (MEXT), Japan.
The work of Y.\,O.~is supported by Grant-in-Aid for Scientific research from the MEXT, Japan, No.\,19K03867.
This work is also supported by 
the Japan Society for the Promotion of Science (JSPS)  Core-to-Core Program, 
No.\,JPJSCCA20200002.


\bibliographystyle{utphys28mod}

\bibliography{ref}

\providecommand{\href}[2]{#2}\begingroup\raggedright\begin{thebibliography}{10}

\bibitem{ATLAS:2021moa}
{\bfseries ATLAS} Collaboration, ``{Search for chargino\textendash{}neutralino
  pair production in final states with three leptons and missing transverse
  momentum in $\sqrt{s} = 13$~TeV pp collisions with the ATLAS detector},''
  \href{https://dx.doi.org/10.1140/epjc/s10052-021-09749-7}{Eur.\  Phys.\  J.\
  C {\bfseries 81} (2021) 1118} {\ttfamily
  [\href{https://arxiv.org/abs/2106.01676}{arXiv:2106.01676}]}.

\bibitem{ATLAS:2022rme}
{\bfseries ATLAS} Collaboration, ``{Search for long-lived charginos based on a
  disappearing-track signature using 136 fb$^{-1}$ of pp collisions at
  $\sqrt{s}$~=~13~TeV with the ATLAS detector},''
  \href{https://dx.doi.org/10.1140/epjc/s10052-022-10489-5}{Eur.\  Phys.\  J.\
  C {\bfseries 82} (2022) 606} {\ttfamily
  [\href{https://arxiv.org/abs/2201.02472}{arXiv:2201.02472}]}.

\bibitem{CMS:2022rbd}
{\bfseries CMS} Collaboration, ``{Searches for additional Higgs bosons and
  vector leptoquarks in $\tau\tau$ final states in proton-proton collisions at
  $\sqrt{s}=13~\mathrm{TeV}$}.'' \url{http://cds.cern.ch/record/2803739}.

\bibitem{CMS:2011lzj}
{\bfseries CMS} Collaboration, ``{Search for Neutral MSSM Higgs Bosons Decaying
  to Tau Pairs in $pp$ Collisions at $\sqrt{s}=7$ TeV},''
  \href{https://dx.doi.org/10.1103/PhysRevLett.106.231801}{Phys.\  Rev.\
  Lett.\  {\bfseries 106} (2011) 231801} {\ttfamily
  [\href{https://arxiv.org/abs/1104.1619}{arXiv:1104.1619}]}.

\bibitem{ATLAS:2012ube}
{\bfseries ATLAS} Collaboration, ``{Search for the neutral Higgs bosons of the
  Minimal Supersymmetric Standard Model in $pp$ collisions at $\sqrt{s}=7$ TeV
  with the ATLAS detector},''
  \href{https://dx.doi.org/10.1007/JHEP02(2013)095}{JHEP {\bfseries 02} (2013)
  095} {\ttfamily [\href{https://arxiv.org/abs/1211.6956}{arXiv:1211.6956}]}.

\bibitem{ATLAS:2014vhc}
{\bfseries ATLAS} Collaboration, ``{Search for neutral Higgs bosons of the
  minimal supersymmetric standard model in pp collisions at $\sqrt{s}$ = 8 TeV
  with the ATLAS detector},''
  \href{https://dx.doi.org/10.1007/JHEP11(2014)056}{JHEP {\bfseries 11} (2014)
  056} {\ttfamily [\href{https://arxiv.org/abs/1409.6064}{arXiv:1409.6064}]}.

\bibitem{ATLAS:2016ivh}
{\bfseries ATLAS} Collaboration, ``{Search for Minimal Supersymmetric Standard
  Model Higgs bosons $H/A$ and for a $Z^{\prime}$ boson in the $\tau \tau$
  final state produced in $pp$ collisions at $\sqrt{s}=13$ TeV with the ATLAS
  Detector},'' \href{https://dx.doi.org/10.1140/epjc/s10052-016-4400-6}{Eur.\
  Phys.\  J.\  C {\bfseries 76} (2016) 585} {\ttfamily
  [\href{https://arxiv.org/abs/1608.00890}{arXiv:1608.00890}]}.

\bibitem{ATLAS:2017eiz}
{\bfseries ATLAS} Collaboration, ``{Search for additional heavy neutral Higgs
  and gauge bosons in the ditau final state produced in 36 fb$^{-1}$ of pp
  collisions at $ \sqrt{s}=13 $ TeV with the ATLAS detector},''
  \href{https://dx.doi.org/10.1007/JHEP01(2018)055}{JHEP {\bfseries 01} (2018)
  055} {\ttfamily [\href{https://arxiv.org/abs/1709.07242}{arXiv:1709.07242}]}.

\bibitem{CMS:2018rmh}
{\bfseries CMS} Collaboration, ``{Search for additional neutral MSSM Higgs
  bosons in the $\tau\tau$ final state in proton-proton collisions at
  $\sqrt{s}=$ 13 TeV},'' \href{https://dx.doi.org/10.1007/JHEP09(2018)007}{JHEP
  {\bfseries 09} (2018) 007} {\ttfamily
  [\href{https://arxiv.org/abs/1803.06553}{arXiv:1803.06553}]}.

\bibitem{ATLAS:2020zms}
{\bfseries ATLAS} Collaboration, ``{Search for heavy Higgs bosons decaying into
  two tau leptons with the ATLAS detector using $pp$ collisions at
  $\sqrt{s}=13$ TeV},''
  \href{https://dx.doi.org/10.1103/PhysRevLett.125.051801}{Phys.\  Rev.\
  Lett.\  {\bfseries 125} (2020) 051801} {\ttfamily
  [\href{https://arxiv.org/abs/2002.12223}{arXiv:2002.12223}]}.

\bibitem{Anastasiou:2016cez}
C.~Anastasiou, {\em et al.}, ``{High precision determination of the gluon
  fusion Higgs boson cross-section at the LHC},''
  \href{https://dx.doi.org/10.1007/JHEP05(2016)058}{JHEP {\bfseries 05} (2016)
  058} {\ttfamily [\href{https://arxiv.org/abs/1602.00695}{arXiv:1602.00695}]}.

\bibitem{CMS:2015ocq}
{\bfseries CMS} Collaboration, ``{Search for new resonances in the diphoton
  final state in the mass range between 80 and 115 GeV in pp collisions at
  $\sqrt{s}=8$ TeV}.'' \url{https://cds.cern.ch/record/2063739}.

\bibitem{CMS:2018cyk}
{\bfseries CMS} Collaboration, ``{Search for a standard model-like Higgs boson
  in the mass range between 70 and 110 GeV in the diphoton final state in
  proton-proton collisions at $\sqrt{s}=$ 8 and 13 TeV},''
  \href{https://dx.doi.org/10.1016/j.physletb.2019.03.064}{Phys.\  Lett.\  B
  {\bfseries 793} (2019) 320--347} {\ttfamily
  [\href{https://arxiv.org/abs/1811.08459}{arXiv:1811.08459}]}.

\bibitem{ATLAS:2018xad}
{\bfseries ATLAS} Collaboration, ``{Search for resonances in the 65 to 110 GeV
  diphoton invariant mass range using 80 fb$^{-1}$ of $pp$ collisions collected
  at $\sqrt{s}=13$ TeV with the ATLAS detector}.''
  \url{https://atlas.web.cern.ch/Atlas/GROUPS/PHYSICS/CONFNOTES/ATLAS-CONF-2018-025/}.

\bibitem{Heinemeyer:2018wzl}
S.~Heinemeyer and T.~Stefaniak, ``{A Higgs Boson at 96 GeV?!}''
  \href{https://dx.doi.org/10.22323/1.339.0016}{PoS {\bfseries CHARGED2018}
  (2019) 016} {\ttfamily
  [\href{https://arxiv.org/abs/1812.05864}{arXiv:1812.05864}]}.

\bibitem{LEPWorkingGroupforHiggsbosonsearches:2003ing}
{\bfseries LEP Working Group for Higgs boson searches, ALEPH, DELPHI, L3, OPAL}
  Collaboration, ``{Search for the standard model Higgs boson at LEP},''
  \href{https://dx.doi.org/10.1016/S0370-2693(03)00614-2}{Phys.\  Lett.\  B
  {\bfseries 565} (2003) 61--75} {\ttfamily
  [\href{https://arxiv.org/abs/hep-ex/0306033}{hep-ex/0306033}]}.

\bibitem{Fox:2017uwr}
P.~J.~Fox and N.~Weiner, ``{Light Signals from a Lighter Higgs},''
  \href{https://dx.doi.org/10.1007/JHEP08(2018)025}{JHEP {\bfseries 08} (2018)
  025} {\ttfamily [\href{https://arxiv.org/abs/1710.07649}{arXiv:1710.07649}]}.

\bibitem{Haisch:2017gql}
U.~Haisch and A.~Malinauskas, ``{Let there be light from a second light Higgs
  doublet},'' \href{https://dx.doi.org/10.1007/JHEP03(2018)135}{JHEP {\bfseries
  03} (2018) 135} {\ttfamily
  [\href{https://arxiv.org/abs/1712.06599}{arXiv:1712.06599}]}.

\bibitem{Biekotter:2017xmf}
T.~Biek\"otter, S.~Heinemeyer, and C.~Mu\~noz, ``{Precise prediction for the
  Higgs-boson masses in the $\mu \nu $ SSM},''
  \href{https://dx.doi.org/10.1140/epjc/s10052-018-5978-7}{Eur.\  Phys.\  J.\
  C {\bfseries 78} (2018) 504} {\ttfamily
  [\href{https://arxiv.org/abs/1712.07475}{arXiv:1712.07475}]}.

\bibitem{Liu:2018xsw}
D.~Liu, J.~Liu, C.~E.~M.~Wagner, and X.-P.~Wang, ``{A Light Higgs at the LHC
  and the B-Anomalies},''
  \href{https://dx.doi.org/10.1007/JHEP06(2018)150}{JHEP {\bfseries 06} (2018)
  150} {\ttfamily [\href{https://arxiv.org/abs/1805.01476}{arXiv:1805.01476}]}.

\bibitem{Domingo:2018uim}
F.~Domingo, S.~Heinemeyer, S.~Pa\ss{}ehr, and G.~Weiglein, ``{Decays of the
  neutral Higgs bosons into SM fermions and gauge bosons in the
  $\mathcal{CP}$-violating NMSSM},''
  \href{https://dx.doi.org/10.1140/epjc/s10052-018-6400-1}{Eur.\  Phys.\  J.\
  C {\bfseries 78} (2018) 942} {\ttfamily
  [\href{https://arxiv.org/abs/1807.06322}{arXiv:1807.06322}]}.

\bibitem{Hollik:2018yek}
W.~G.~Hollik, S.~Liebler, G.~Moortgat-Pick, S.~Pa\ss{}ehr, and G.~Weiglein,
  ``{Phenomenology of the inflation-inspired NMSSM at the electroweak scale},''
  \href{https://dx.doi.org/10.1140/epjc/s10052-019-6561-6}{Eur.\  Phys.\  J.\
  C {\bfseries 79} (2019) 75} {\ttfamily
  [\href{https://arxiv.org/abs/1809.07371}{arXiv:1809.07371}]}.

\bibitem{Biekotter:2019kde}
T.~Biek\"otter, M.~Chakraborti, and S.~Heinemeyer, ``{A 96 GeV Higgs boson in
  the N2HDM},'' \href{https://dx.doi.org/10.1140/epjc/s10052-019-7561-2}{Eur.\
  Phys.\  J.\  C {\bfseries 80} (2020) 2} {\ttfamily
  [\href{https://arxiv.org/abs/1903.11661}{arXiv:1903.11661}]}.

\bibitem{Cline:2019okt}
J.~M.~Cline and T.~Toma, ``{Pseudo-Goldstone dark matter confronts cosmic ray
  and collider anomalies},''
  \href{https://dx.doi.org/10.1103/PhysRevD.100.035023}{Phys.\  Rev.\  D
  {\bfseries 100} (2019) 035023} {\ttfamily
  [\href{https://arxiv.org/abs/1906.02175}{arXiv:1906.02175}]}.

\bibitem{Cao:2019ofo}
J.~Cao, X.~Jia, Y.~Yue, H.~Zhou, and P.~Zhu, ``{96 GeV diphoton excess in
  seesaw extensions of the natural NMSSM},''
  \href{https://dx.doi.org/10.1103/PhysRevD.101.055008}{Phys.\  Rev.\  D
  {\bfseries 101} (2020) 055008} {\ttfamily
  [\href{https://arxiv.org/abs/1908.07206}{arXiv:1908.07206}]}.

\bibitem{Aguilar-Saavedra:2020wrj}
J.~A.~Aguilar-Saavedra and F.~R.~Joaquim, ``{Multiphoton signals of a (96 GeV?)
  stealth boson},''
  \href{https://dx.doi.org/10.1140/epjc/s10052-020-7952-4}{Eur.\  Phys.\  J.\
  C {\bfseries 80} (2020) 403} {\ttfamily
  [\href{https://arxiv.org/abs/2002.07697}{arXiv:2002.07697}]}.

\bibitem{Biekotter:2020cjs}
T.~Biek\"otter, M.~Chakraborti, and S.~Heinemeyer, ``{The \textquotedblleft{}96
  GeV excess\textquotedblright{} at the LHC},''
  \href{https://dx.doi.org/10.1142/S0217751X21420185}{Int.\  J.\  Mod.\  Phys.\
   A {\bfseries 36} (2021) 2142018} {\ttfamily
  [\href{https://arxiv.org/abs/2003.05422}{arXiv:2003.05422}]}.

\bibitem{Biekotter:2021ovi}
T.~Biek\"otter and M.~O.~Olea-Romacho, ``{Reconciling Higgs physics and
  pseudo-Nambu-Goldstone dark matter in the S2HDM using a genetic algorithm},''
  \href{https://dx.doi.org/10.1007/JHEP10(2021)215}{JHEP {\bfseries 10} (2021)
  215} {\ttfamily [\href{https://arxiv.org/abs/2108.10864}{arXiv:2108.10864}]}.

\bibitem{Biekotter:2021qbc}
T.~Biek\"otter, A.~Grohsjean, S.~Heinemeyer, C.~Schwanenberger, and
  G.~Weiglein, ``{Possible indications for new Higgs bosons in the reach of the
  LHC: N2HDM and NMSSM interpretations},''
  \href{https://dx.doi.org/10.1140/epjc/s10052-022-10099-1}{Eur.\  Phys.\  J.\
  C {\bfseries 82} (2022) 178} {\ttfamily
  [\href{https://arxiv.org/abs/2109.01128}{arXiv:2109.01128}]}.

\bibitem{Heinemeyer:2021msz}
S.~Heinemeyer, C.~Li, F.~Lika, G.~Moortgat-Pick, and S.~Paasch, ``{A 96 GeV
  Higgs Boson in the 2HDM plus Singlet}.'' {\ttfamily
  \href{https://arxiv.org/abs/2112.11958}{arXiv:2112.11958}}.

\bibitem{Benbrik:2022azi}
R.~Benbrik, M.~Boukidi, S.~Moretti, and S.~Semlali, ``{Explaining the 96 GeV
  Di-photon anomaly in a generic 2HDM Type-III},''
  \href{https://dx.doi.org/10.1016/j.physletb.2022.137245}{Phys.\  Lett.\  B
  {\bfseries 832} (2022) 137245} {\ttfamily
  [\href{https://arxiv.org/abs/2204.07470}{arXiv:2204.07470}]}.

\bibitem{Benbrik:2022dja}
R.~Benbrik, M.~Boukidi, and B.~Manaut, ``{$W$-mass and 96 GeV excess in
  type-III 2HDM}.'' {\ttfamily
  \href{https://arxiv.org/abs/2204.11755}{arXiv:2204.11755}}.

\bibitem{Biekotter:2022jyr}
T.~Biek\"otter, S.~Heinemeyer, and G.~Weiglein, ``{Mounting evidence for a 95
  GeV Higgs boson},'' \href{https://dx.doi.org/10.1007/JHEP08(2022)201}{JHEP
  {\bfseries 08} (2022) 201} {\ttfamily
  [\href{https://arxiv.org/abs/2203.13180}{arXiv:2203.13180}]}.

\bibitem{Biekotter:2022abc}
T.~Biek\"otter, S.~Heinemeyer, and G.~Weiglein, ``{Excesses in the low-mass
  Higgs-boson search and the W-boson mass measurement}.'' {\ttfamily
  \href{https://arxiv.org/abs/2204.05975}{arXiv:2204.05975}}.

\bibitem{Kumar:2012ww}
K.~Kumar, R.~Vega-Morales, and F.~Yu, ``{Effects from New Colored States and
  the Higgs Portal on Gluon Fusion and Higgs Decays},''
  \href{https://dx.doi.org/10.1103/PhysRevD.86.113002}{Phys.\  Rev.\  D
  {\bfseries 86} (2012) 113002} {\ttfamily
  [\href{https://arxiv.org/abs/1205.4244}{arXiv:1205.4244}]}. [Erratum:
  Phys.Rev.D 87, 119903 (2013)].

\bibitem{Abdullah:2019dpu}
M.~Abdullah, M.~Dalchenko, T.~Kamon, D.~Rathjens, and A.~Thompson, ``{A heavy
  neutral gauge boson near the $Z$ boson mass pole via third generation
  fermions at the LHC},''
  \href{https://dx.doi.org/10.1016/j.physletb.2020.135326}{Phys.\  Lett.\  B
  {\bfseries 803} (2020) 135326} {\ttfamily
  [\href{https://arxiv.org/abs/1912.00102}{arXiv:1912.00102}]}.

\bibitem{Campbell:2011cu}
J.~M.~Campbell, R.~K.~Ellis, and C.~Williams, ``{Gluon-Gluon Contributions to
  W+ W- Production and Higgs Interference Effects},''
  \href{https://dx.doi.org/10.1007/JHEP10(2011)005}{JHEP {\bfseries 10} (2011)
  005} {\ttfamily [\href{https://arxiv.org/abs/1107.5569}{arXiv:1107.5569}]}.

\bibitem{Pich:2009sp}
A.~Pich and P.~Tuzon, ``{Yukawa Alignment in the Two-Higgs-Doublet Model},''
  \href{https://dx.doi.org/10.1103/PhysRevD.80.091702}{Phys.\  Rev.\  D
  {\bfseries 80} (2009) 091702} {\ttfamily
  [\href{https://arxiv.org/abs/0908.1554}{arXiv:0908.1554}]}.

\bibitem{Branco:2011iw}
G.~C.~Branco, {\em et al.}, ``{Theory and phenomenology of two-Higgs-doublet
  models},'' \href{https://dx.doi.org/10.1016/j.physrep.2012.02.002}{Phys.\
  Rept.\  {\bfseries 516} (2012) 1--102} {\ttfamily
  [\href{https://arxiv.org/abs/1106.0034}{arXiv:1106.0034}]}.

\bibitem{Cline:2015lqp}
J.~M.~Cline, ``{Scalar doublet models confront $\tau$ and $b$ anomalies},''
  \href{https://dx.doi.org/10.1103/PhysRevD.93.075017}{Phys.\  Rev.\  D
  {\bfseries 93} (2016) 075017} {\ttfamily
  [\href{https://arxiv.org/abs/1512.02210}{arXiv:1512.02210}]}.

\bibitem{Georgi:1986df}
H.~Georgi, D.~B.~Kaplan, and L.~Randall, ``{Manifesting the Invisible Axion at
  Low-energies},''
  \href{https://dx.doi.org/10.1016/0370-2693(86)90688-X}{Phys.\  Lett.\  B
  {\bfseries 169} (1986) 73--78}.

\bibitem{Ellwanger:2009dp}
U.~Ellwanger, C.~Hugonie, and A.~M.~Teixeira, ``{The Next-to-Minimal
  Supersymmetric Standard Model},''
  \href{https://dx.doi.org/10.1016/j.physrep.2010.07.001}{Phys.\  Rept.\
  {\bfseries 496} (2010) 1--77} {\ttfamily
  [\href{https://arxiv.org/abs/0910.1785}{arXiv:0910.1785}]}.

\bibitem{ATLAS:2018gfm}
{\bfseries ATLAS} Collaboration, ``{Search for charged Higgs bosons decaying
  via $H^{\pm} \to \tau^{\pm}\nu_{\tau}$ in the $\tau$+jets and $\tau$+lepton
  final states with 36 fb$^{-1}$ of $pp$ collision data recorded at $\sqrt{s} =
  13$ TeV with the ATLAS experiment},''
  \href{https://dx.doi.org/10.1007/JHEP09(2018)139}{JHEP {\bfseries 09} (2018)
  139} {\ttfamily [\href{https://arxiv.org/abs/1807.07915}{arXiv:1807.07915}]}.

\bibitem{ATLAS:2021upq}
{\bfseries ATLAS} Collaboration, ``{Search for charged Higgs bosons decaying
  into a top quark and a bottom quark at $ \sqrt{\mathrm{s}} $ = 13 TeV with
  the ATLAS detector},'' \href{https://dx.doi.org/10.1007/JHEP06(2021)145}{JHEP
  {\bfseries 06} (2021) 145} {\ttfamily
  [\href{https://arxiv.org/abs/2102.10076}{arXiv:2102.10076}]}.

\bibitem{Harlander:2012pb}
R.~V.~Harlander, S.~Liebler, and H.~Mantler, ``{SusHi: A program for the
  calculation of Higgs production in gluon fusion and bottom-quark annihilation
  in the Standard Model and the MSSM},''
  \href{https://dx.doi.org/10.1016/j.cpc.2013.02.006}{Comput.\  Phys.\
  Commun.\  {\bfseries 184} (2013) 1605--1617} {\ttfamily
  [\href{https://arxiv.org/abs/1212.3249}{arXiv:1212.3249}]}.

\bibitem{Harlander:2016hcx}
R.~V.~Harlander, S.~Liebler, and H.~Mantler, ``{SusHi Bento: Beyond NNLO and
  the heavy-top limit},''
  \href{https://dx.doi.org/10.1016/j.cpc.2016.10.015}{Comput.\  Phys.\
  Commun.\  {\bfseries 212} (2017) 239--257} {\ttfamily
  [\href{https://arxiv.org/abs/1605.03190}{arXiv:1605.03190}]}.

\bibitem{Alwall:2014hca}
J.~Alwall, {\em et al.}, ``{The automated computation of tree-level and
  next-to-leading order differential cross sections, and their matching to
  parton shower simulations},''
  \href{https://dx.doi.org/10.1007/JHEP07(2014)079}{JHEP {\bfseries 07} (2014)
  079} {\ttfamily [\href{https://arxiv.org/abs/1405.0301}{arXiv:1405.0301}]}.

\bibitem{Frixione:2014qaa}
S.~Frixione, V.~Hirschi, D.~Pagani, H.~S.~Shao, and M.~Zaro, ``{Weak
  corrections to Higgs hadroproduction in association with a top-quark pair},''
  \href{https://dx.doi.org/10.1007/JHEP09(2014)065}{JHEP {\bfseries 09} (2014)
  065} {\ttfamily [\href{https://arxiv.org/abs/1407.0823}{arXiv:1407.0823}]}.

\bibitem{ATLAS:2018mme}
{\bfseries ATLAS} Collaboration, ``{Observation of Higgs boson production in
  association with a top quark pair at the LHC with the ATLAS detector},''
  \href{https://dx.doi.org/10.1016/j.physletb.2018.07.035}{Phys.\  Lett.\  B
  {\bfseries 784} (2018) 173--191} {\ttfamily
  [\href{https://arxiv.org/abs/1806.00425}{arXiv:1806.00425}]}.

\bibitem{ATLAS:2021fzm}
{\bfseries ATLAS} Collaboration, ``{Measurements of the inclusive and
  differential production cross sections of a top-quark\textendash{}antiquark
  pair in association with a Z~boson at $\sqrt{s} = 13$~TeV with the ATLAS
  detector},'' \href{https://dx.doi.org/10.1140/epjc/s10052-021-09439-4}{Eur.\
  Phys.\  J.\  C {\bfseries 81} (2021) 737} {\ttfamily
  [\href{https://arxiv.org/abs/2103.12603}{arXiv:2103.12603}]}.

\bibitem{Frixione:2015zaa}
S.~Frixione, V.~Hirschi, D.~Pagani, H.~S.~Shao, and M.~Zaro, ``{Electroweak and
  QCD corrections to top-pair hadroproduction in association with heavy
  bosons},'' \href{https://dx.doi.org/10.1007/JHEP06(2015)184}{JHEP {\bfseries
  06} (2015) 184} {\ttfamily
  [\href{https://arxiv.org/abs/1504.03446}{arXiv:1504.03446}]}.

\bibitem{Frederix:2011zi}
R.~Frederix, {\em et al.}, ``{Scalar and pseudoscalar Higgs production in
  association with a top\textendash{}antitop pair},''
  \href{https://dx.doi.org/10.1016/j.physletb.2011.06.012}{Phys.\  Lett.\  B
  {\bfseries 701} (2011) 427--433} {\ttfamily
  [\href{https://arxiv.org/abs/1104.5613}{arXiv:1104.5613}]}.

\bibitem{Djouadi:2005gj}
A.~Djouadi, ``{The Anatomy of electro-weak symmetry breaking. II. The Higgs
  bosons in the minimal supersymmetric model},''
  \href{https://dx.doi.org/10.1016/j.physrep.2007.10.005}{Phys.\  Rept.\
  {\bfseries 459} (2008) 1--241} {\ttfamily
  [\href{https://arxiv.org/abs/hep-ph/0503173}{hep-ph/0503173}]}.

\bibitem{Dolan:2016qvg}
M.~J.~Dolan, M.~Spannowsky, Q.~Wang, and Z.-H.~Yu, ``{Determining the quantum
  numbers of simplified models in $t\bar{t}X$ production at the LHC},''
  \href{https://dx.doi.org/10.1103/PhysRevD.94.015025}{Phys.\  Rev.\  D
  {\bfseries 94} (2016) 015025} {\ttfamily
  [\href{https://arxiv.org/abs/1606.00019}{arXiv:1606.00019}]}.

\bibitem{ATLAS:2022yrq}
{\bfseries ATLAS} Collaboration, ``{Measurements of Higgs boson production
  cross-sections in the $H\to\tau^{+}\tau^{-}$ decay channel in $pp$ collisions
  at $\sqrt{s}=13\,\text{TeV}$ with the ATLAS detector},''
  \href{https://dx.doi.org/10.1007/JHEP08(2022)175}{JHEP {\bfseries 08} (2022)
  175} {\ttfamily [\href{https://arxiv.org/abs/2201.08269}{arXiv:2201.08269}]},
  \url{https://www.hepdata.net/record/ins2014187}.

\bibitem{Gunion:1989we}
J.~F.~Gunion, H.~E.~Haber, G.~L.~Kane, and S.~Dawson, {\em {The Higgs Hunter's
  Guide}}, vol.~80.
\newblock 2000.

\bibitem{Denner:2011mq}
A.~Denner, S.~Heinemeyer, I.~Puljak, D.~Rebuzzi, and M.~Spira, ``{Standard
  Model Higgs-Boson Branching Ratios with Uncertainties},''
  \href{https://dx.doi.org/10.1140/epjc/s10052-011-1753-8}{Eur.\  Phys.\  J.\
  C {\bfseries 71} (2011) 1753} {\ttfamily
  [\href{https://arxiv.org/abs/1107.5909}{arXiv:1107.5909}]}.

\bibitem{Landau:1948kw}
L.~D.~Landau, ``{On the angular momentum of a system of two photons},''
\href{https://dx.doi.org/10.1016/B978-0-08-010586-4.50070-5}{Dokl.\  Akad.\
  Nauk Ser.\  Fiz.\  {\bfseries 60} (1948) 207--209}.

\bibitem{Yang:1950rg}
C.-N.~Yang, ``{Selection Rules for the Dematerialization of a Particle Into Two
  Photons},''
\href{https://dx.doi.org/10.1103/PhysRev.77.242}{Phys.\  Rev.\  {\bfseries 77}
  (1950) 242--245}.

\bibitem{ATLAS:2020ior}
{\bfseries ATLAS} Collaboration, ``{$CP$ Properties of Higgs Boson Interactions
  with Top Quarks in the $t\bar{t}H$ and $tH$ Processes Using $H \rightarrow
  \gamma\gamma$ with the ATLAS Detector},''
  \href{https://dx.doi.org/10.1103/PhysRevLett.125.061802}{Phys.\  Rev.\
  Lett.\  {\bfseries 125} (2020) 061802} {\ttfamily
  [\href{https://arxiv.org/abs/2004.04545}{arXiv:2004.04545}]}.

\bibitem{CMS:2021kom}
{\bfseries CMS} Collaboration, ``{Measurements of Higgs boson production cross
  sections and couplings in the diphoton decay channel at $ \sqrt{\mathrm{s}} $
  = 13 TeV},'' \href{https://dx.doi.org/10.1007/JHEP07(2021)027}{JHEP
  {\bfseries 07} (2021) 027} {\ttfamily
  [\href{https://arxiv.org/abs/2103.06956}{arXiv:2103.06956}]}.

\bibitem{ATLAS:2017ztq}
{\bfseries ATLAS} Collaboration, ``{Evidence for the associated production of
  the Higgs boson and a top quark pair with the ATLAS detector},''
  \href{https://dx.doi.org/10.1103/PhysRevD.97.072003}{Phys.\  Rev.\  D
  {\bfseries 97} (2018) 072003} {\ttfamily
  [\href{https://arxiv.org/abs/1712.08891}{arXiv:1712.08891}]}.

\bibitem{ATLAS:2019nvo}
{\bfseries ATLAS} Collaboration, ``{Analysis of $t\bar{t}H$ and $t\bar{t}W$
  production in multilepton final states with the ATLAS detector}.''
  \url{https://cds.cern.ch/record/2693930}.

\bibitem{Fuyuto:2015ida}
K.~Fuyuto, J.~Hisano, and E.~Senaha, ``{Toward verification of electroweak
  baryogenesis by electric dipole moments},''
  \href{https://dx.doi.org/10.1016/j.physletb.2016.02.053}{Phys.\  Lett.\  B
  {\bfseries 755} (2016) 491--497} {\ttfamily
  [\href{https://arxiv.org/abs/1510.04485}{arXiv:1510.04485}]}.

\bibitem{ATLAS:2017fak}
{\bfseries ATLAS} Collaboration, ``{Search for the standard model Higgs boson
  produced in association with top quarks and decaying into a $b\bar{b}$ pair
  in $pp$ collisions at $\sqrt{s}$ = 13 TeV with the ATLAS detector},''
  \href{https://dx.doi.org/10.1103/PhysRevD.97.072016}{Phys.\  Rev.\  D
  {\bfseries 97} (2018) 072016} {\ttfamily
  [\href{https://arxiv.org/abs/1712.08895}{arXiv:1712.08895}]}.

\bibitem{Fayet:2007ua}
P.~Fayet, ``{U-boson production in e+ e- annihilations, psi and Upsilon decays,
  and Light Dark Matter},''
  \href{https://dx.doi.org/10.1103/PhysRevD.75.115017}{Phys.\  Rev.\  D
  {\bfseries 75} (2007) 115017} {\ttfamily
  [\href{https://arxiv.org/abs/hep-ph/0702176}{hep-ph/0702176}]}.

\bibitem{Pospelov:2008zw}
M.~Pospelov, ``{Secluded U(1) below the weak scale},''
  \href{https://dx.doi.org/10.1103/PhysRevD.80.095002}{Phys.\  Rev.\  D
  {\bfseries 80} (2009) 095002} {\ttfamily
  [\href{https://arxiv.org/abs/0811.1030}{arXiv:0811.1030}]}.

\bibitem{Buschmann:2015awa}
M.~Buschmann, J.~Kopp, J.~Liu, and P.~A.~N.~Machado, ``{Lepton Jets from
  Radiating Dark Matter},''
  \href{https://dx.doi.org/10.1007/JHEP07(2015)045}{JHEP {\bfseries 07} (2015)
  045} {\ttfamily [\href{https://arxiv.org/abs/1505.07459}{arXiv:1505.07459}]}.

\bibitem{CMS:2019buh}
{\bfseries CMS} Collaboration, ``{Search for a Narrow Resonance Lighter than
  200 GeV Decaying to a Pair of Muons in Proton-Proton Collisions at $\sqrt{s}
  =$ TeV},'' \href{https://dx.doi.org/10.1103/PhysRevLett.124.131802}{Phys.\
  Rev.\  Lett.\  {\bfseries 124} (2020) 131802} {\ttfamily
  [\href{https://arxiv.org/abs/1912.04776}{arXiv:1912.04776}]}.

\bibitem{Curtin:2014cca}
D.~Curtin, R.~Essig, S.~Gori, and J.~Shelton, ``{Illuminating Dark Photons with
  High-Energy Colliders},''
  \href{https://dx.doi.org/10.1007/JHEP02(2015)157}{JHEP {\bfseries 02} (2015)
  157} {\ttfamily [\href{https://arxiv.org/abs/1412.0018}{arXiv:1412.0018}]}.

\bibitem{Iguro:2017ysu}
S.~Iguro and K.~Tobe, ``{$R(D^{(*)})$ in a general two Higgs doublet model},''
  \href{https://dx.doi.org/10.1016/j.nuclphysb.2017.10.014}{Nucl.\  Phys.\  B
  {\bfseries 925} (2017) 560--606} {\ttfamily
  [\href{https://arxiv.org/abs/1708.06176}{arXiv:1708.06176}]}.

\bibitem{Iguro:2022uzz}
S.~Iguro, ``{Revival of H- interpretation of RD(*) anomaly and closing low mass
  window},'' \href{https://dx.doi.org/10.1103/PhysRevD.105.095011}{Phys.\
  Rev.\  D {\bfseries 105} (2022) 095011} {\ttfamily
  [\href{https://arxiv.org/abs/2201.06565}{arXiv:2201.06565}]}.

\bibitem{Aoki:2021kgd}
Y.~Aoki {\em et~al.}, ``{FLAG Review 2021}.'' {\ttfamily
  \href{https://arxiv.org/abs/2111.09849}{arXiv:2111.09849}}.

\bibitem{Iguro:2020cpg}
S.~Iguro and R.~Watanabe, ``{Bayesian fit analysis to full distribution data of
  $ \overline{\mathrm{B}}\to {\mathrm{D}}^{\left(\ast
  \right)}\mathrm{\ell}\overline{\nu }:\left|{\mathrm{V}}_{\mathrm{cb}}\right|
  $ determination and new physics constraints},''
  \href{https://dx.doi.org/10.1007/JHEP08(2020)006}{JHEP {\bfseries 08} (2020)
  006} {\ttfamily [\href{https://arxiv.org/abs/2004.10208}{arXiv:2004.10208}]}.

\bibitem{Blanke:2022pjy}
M.~Blanke, S.~Iguro, and H.~Zhang, ``{Towards ruling out the charged Higgs
  interpretation of the $ {R}_{D^{\left(\ast \right)}} $ anomaly},''
  \href{https://dx.doi.org/10.1007/JHEP06(2022)043}{JHEP {\bfseries 06} (2022)
  043} {\ttfamily [\href{https://arxiv.org/abs/2202.10468}{arXiv:2202.10468}]}.

\bibitem{Iguro:2018fni}
S.~Iguro, Y.~Omura, and M.~Takeuchi, ``{Test of the $R(D^{(*)})$ anomaly at the
  LHC},'' \href{https://dx.doi.org/10.1103/PhysRevD.99.075013}{Phys.\  Rev.\  D
  {\bfseries 99} (2019) 075013} {\ttfamily
  [\href{https://arxiv.org/abs/1810.05843}{arXiv:1810.05843}]}.

\bibitem{Iguro:2018qzf}
S.~Iguro and Y.~Omura, ``{Status of the semileptonic $B$ decays and muon g-2 in
  general 2HDMs with right-handed neutrinos},''
  \href{https://dx.doi.org/10.1007/JHEP05(2018)173}{JHEP {\bfseries 05} (2018)
  173} {\ttfamily [\href{https://arxiv.org/abs/1802.01732}{arXiv:1802.01732}]}.

\bibitem{ATLAS:2021fet}
{\bfseries ATLAS} Collaboration, ``{Search for resonant and non-resonant Higgs
  boson pair production in the $b\bar b\tau^+\tau^-$ decay channel using 13 TeV
  $pp$ collision data from the ATLAS detector}.''
  \url{http://cds.cern.ch/record/2777236}.

\end{thebibliography}\endgroup

\end{document}